\documentclass{aa}
\usepackage{txfonts,textcomp}
\usepackage{graphicx}
\usepackage{natbib}
\def\arcsec{\hbox{$^{\prime\prime}$}}

\def\arcsec{\hbox{$^{\prime\prime}$}}

\usepackage{multirow}
\usepackage{makecell}

\begin{document}

\title{Multi-purpose InSTRument for Astronomy at Low-resolution: MISTRAL@OHP
 ~\thanks{Based on observations obtained with Observatoire de Haute-Provence (OHP) instruments  
(see acknowledgements for more details).}}

\author{
J.~Schmitt\inst{1} \and
C.~Adami\inst{2} \and
M.~Dennefeld\inst{3} \and \\
F.~Agneray\inst{2} \and
S.~Basa\inst{1,2} \and
J.C.~Brunel\inst{1} \and
V.~Buat\inst{2} \and
D.~Burgarella\inst{2} \and
C.~Carvalho\inst{3} \and
G.~Castagnoli\inst{1} \and
N.~Grosso\inst{2} \and
F.~Huppert\inst{1} \and
C.~Moreau\inst{2} \and
F.~Moreau\inst{1} \and
L.~Moreau\inst{1} \and
E.~Muslimov\inst{2} \and
S.~Pascal\inst{1} \and
S.~Perruchot\inst{1} \and
D.~Russeil\inst{2}\and \\
J.L.~Beuzit\inst{4} \and
F.~Dolon\inst{1} \and
M.~Ferrari\inst{1,2} \and
B.~Hamelin\inst{4} \and
A.~LevanSuu\inst{1} \and \\
K.~Aravind\inst{5} \and
D.~Gotz\inst{7} \and
E.~Jehin\inst{6} \and
E.~LeFloc'h\inst{7} \and
J.~Palmerio\inst{8} \and
A.~Saccardi\inst{8} \and
B.~Schneider\inst{7,9} \and
F.~Sch\"ussler\inst{10} \and
D.~Turpin\inst{7} \and
S.D.~Vergani\inst{8}
}

\offprints{C. Adami \email{christophe.adami@lam.fr}}

\institute{
OHP, OSU - Institut Pyth\'eas, UAR 3470, CNRS, Aix-Marseille Universit\'e, 1912 Route de l'Observatoire, 04870 St.Michel l'Observatoire, France
\and
Aix Marseille Univ, CNRS, CNES, LAM, Marseille, France
\and
Sorbonne Université, CNRS, UMR 7095, Institut d’Astrophysique de Paris, 98bis Bd Arago, 75014 Paris, France
\and
OSU - Institut Pyth\'eas, UAR 3470, CNRS, Aix-Marseille Universit\'e, Pôle de l’Étoile Site de Château-Gombert
38, rue Frédéric Joliot-Curie
13388 Marseille, France
\and
Physical Research Laboratory, Navarangpura, Ahmedabad, 380058, Gujarat, India
\and
Space sciences, Technologies \& Astrophysics Research (STAR) Institute, Allée du Six Août, 19C, University of Liège, 4000 Li\'ege, Belgium
\and
CEA, IRFU, DAp, AIM, Université Paris-Saclay, Université Paris Cité, Sorbonne Paris Cité, CNRS, F-91191 Gif-sur-Yvette, France
\and
GEPI, Observatoire de Paris, Université PSL, CNRS, 5 Place Jules Janssen, 92190 Meudon, France
\and
Kavli Institute for Astrophysics and Space Research, Massachusetts Institute of Technology, Cambridge, MA, USA
\and
IRFU, CEA, Université Paris-Saclay, Gif-sur-Yvette, France}

\date{Accepted . Received ; Draft printed: \today}

\authorrunning{Schmitt et al.}

\titlerunning{Multi-purpose InSTRument for Astronomy at Low-resolution: MISTRAL}

\abstract 
% context heading (optional) 
{MISTRAL is the new Faint Object Spectroscopic Camera mounted at the folded Cassegrain focus of the 1.93m telescope of Haute-Provence Observatory. }
% aims heading (mandatory)
{ We describe the design and components of the instrument and  give some details about its operation. }
% methods heading (mandatory)
{ We emphasise in particular the various observing modes and the performances of the detector. A short description is also given about the working environment. Various types of objects, including stars, nebulae, comets, novae, galaxies have been observed during various test phases to evaluate the performances of the instrument.}
% results heading (mandatory)
{ The instrument covers the range of 4000 to 8000~$\AA$ with the blue setting, or from 6000 to 10000~$\AA$ with the red setting, at an average spectral resolution of 700. Its peak efficiency is about 22$\%$ at 6000~$\AA$. In spectroscopy, a limiting magnitude of r$\sim$19.5 can be achieved for a point source in one hour with a signal to noise of 3 in the continuum (and better if emission lines are present).  In imaging mode, limiting magnitudes of 20-21 can be obtained in 10-20mn (with average seing conditions of 2.5 arcsec at OHP). The instrument is very users-friendly and can be put into operations in less than 15mn (rapid change-over from the other instrument in use) if required by the science (like for Gamma-Rays Bursts). 
Some first scientific results are described for various types of objects, and in particular for the follow-up of GRBs.}
% Conclusions (mandatory)
{ While some further improvements are still under way, in particular to ease the switch from blue to red setting and add more grisms or filters, MISTRAL is ready for the follow-up of transients and other variable objects, in the soon-to-come era of e.g. the SVOM satellite and of the Rubin telescope. }


\maketitle

\section{Introduction}
\label{sec:intro}

Since the early 2010th, with the advent of many new sky surveys, both from the ground and from space, the exploration of the variable sky is becoming a highly competitive 
 new domain of Astrophysics (e.g. \cite{2019PASP..131g8001G}, and references therein). The high cadence of those surveys in the framework of Multi-messenger astrophysics, and the large area covered by them 
leads to the discovery of  a wealth of new phenomena and classes of objects, enlarging the observed physical diversity, as well as improving the statistics on previously known types of objects. This will still increase in the near future, with the advent of the Rubin Telescope (former LSST) and the launch of SVOM, thus  requiring a major effort of ground-based follow-up. 
On the high-energy side, Gamma-Rays bursts (GRBs, e.g. \cite{2009ARA&A..47..567G}) are observed   in large numbers, and classified into two categories, the short- and long-duration
GRBs. On the Supernovae (SNe) side (e.g. \cite{2017hsn..book..195G}, it appears that the early classification in types I or II  explosions needs to be refined, to take into account the variety of observed phenomena:  not just core-collapse, or thermonuclear explosions of CO white dwarfs, but including now   ultraluminous  SNe or faint and fast decaying type I SNe,  He detonations Ia objects or objects interacting with the circum-stellar medium (CSM), not forgetting   the variety of  novae. 
The range of underlying physical mechanisms is  therefore much more diverse than previously thought, but is still not understood. On a somewhat quieter
side, Luminous Blue Variables, or the numerous peculiar binaries await a better understanding too.

Non transient, but variable objects also  require some follow-up.  For instance, it has been recently found that some quasars are showing dramatic changes in luminosity and spectral characteristics, the so-called Changing Look Quasars (e.g. \cite{2016MNRAS.457..389M} and references therein), whose changes are not compatible with the Standard Active Galactic Nuclei (AGN) model: more and more objects are found of that kind, and they require more long-term monitoring, to measure timescales and possible delays between the changes in the optical and in other wavelengths. 

On the other hand, large spectroscopic catalogs (e.g. SDSS\footnote{\url{https://skyserver.sdss.org/dr18}}) are rarely  100$\%$ complete or are not always  covering the high latitude regions (see e.g. the XCLASS survey, \cite{2021A&A...652A..12K}). Moreover, spectra available in the literature  are sometimes of poor quality, as they were just made to get a redshift or a  general classification of the objects, with insufficient  signal-to-noise ratio to detect faint characteristic lines, or too short a spectral coverage to e.g. explore the red part of the spectrum.  
What is most needed to progress in all these fields is enough ground-based observing time to follow  over time the evolution of representative examples of all those categories. Both  photometry, and, even more,  spectroscopy is required, including, if possible, the near infrared: only with long time series of  data is it possible to understand the underlying physical mechanisms. Small to medium sized telescopes are now more available
for long time series  than the  8m ones, and are well suited,  provided they are equipped with efficient versatile spectro-imagers.  Observatoire de Haute Provence (OHP) had been pionneer  in the early exploration of the far-red range (6000-11000~$\AA$), first  with the Roucass spectrograph (\cite{1975A&A....41...99A}), and later with the Carelec spectrograph (\cite{1990A&A...228..546L}), but since  the decommission of the latter in 2012, no low dispersion spectroscopic facility was available anymore, at a moment where it is much needed again. 

It is with this in mind that the  MISTRAL\footnote{\url{http://www.obs-hp.fr/guide/mistral/MISTRAL_spectrograph_camera.shtml}} (Multi-purpose InSTRument for Astronomy at Low-dispersion) instrument had been planned  for the OHP 1.93m telescope, with a possibility of rapid reaction (Targets of Opportunity) and changeover   from the other OHP T193 
 instrument, SOPHIE.  While the concept of the focal reducer was invented by G. Courtes in Marseille (\cite{1960AnAp...23..115C}) long ago,   no such instrument was permanently installed at OHP. With the development of Transient Astronomy, such a flexible instrument became mandatory and was envisaged already for the launch of Gaia Alerts. Various issues delayed the project (see below), but  the gap is now filled in, and the MISTRAL instrument entered into service in 2021.

MISTRAL can be operated  in two  modes: regular observing runs in visitor mode and Target of Opportunity (ToO) mode with  service observing  for fast transients. 
MISTRAL can be accessed by any astronomer working in a french astronomical institution via the national call for observing proposals, which uses the NorthStar\footnote{\url{https://northstar.omp.eu/}}  system. Non-nationals can also  access it through the Transnational Access Program of the Opticon RadioNet Pilot program\footnote{\url{https://www.orp-h2020.eu/}}, which includes a wide range of worldwide institutions\footnote{\url{https://www.orp-h2020.eu/partner}}.

In the next section, we describe the MISTRAL main components,  give some details on the operation mode and tools available for the observer.
As on-sky instrument performance validations, we present in section \ref{OSV} real observations demonstrating  MISTRAL's ability to access faint objects of various types, including access to near infrared lines (such as the Paschen lines), and to spectroscopically classify them. To assess MISTRAL's ability to obtain  higher spectral resolution spectra, we also investigate  MISTRAL's spectral resolution around H$\alpha$. This section finally shows that we can follow moving targets with a report of a comet follow-up.
Section \ref{GRBR} reports about the core of  MISTRAL's activities: GRB follow-ups. We detail in this section several observations with an intrinsic scientific interest (e.g. the "BOAT").
Finally, more details about the instrument operations are given in the annexes.

\section{General instrument description} 

\subsection {General set-up}  

A versatile spectro-imager should cover as large a wavelength range as possible, to cover the various science goals described above. But this wish is limited by the size of the detector, here a single one with 2048 pixels. While the initial idea was then to follow the path of SPRAT at the Liverpool telescope (\cite{2017PhDT.......214P}), to speed up the development, it rapidly became clear that several factors were preventing this simple approach. First of all, the 1.93m telescope being a relatively old telescope, its f/ratio at Cassegrain focus is f/15, contrary to more modern telescopes which are at f/10 or faster: to avoid too long an instrument, a focal reducer had thus to be introduced. Second, since the decommissioning of the Carelec, the main instrument at the 1.93m was a high-resolution spectrograph for exoplanets research with the velocimetry technique (SOPHIE), requiring a good long-term stability. Rapid change-overs needed for ToO observations were therefore not compatible with this requirement. Of course, the ideal solution would have been a common adapter for both instruments (SOPHIE being anyway fed by fibers) but this could not be implemented in time. It was therefore decided to mount MISTRAL at the folded Cassegrain focus, the change-over from one to the other instrument being done simply by putting/removing a flat 45$^{\circ}$ mirror in the main beam: the change-over is done this way in less than 15 minutes (including adjustment of the telescope focus). While having then anyway to change the mechanical design with respect to SPRAT, we took advantage of those changes to implement more flexibility in the instrument, like introducing a filter holder  and a grism holder, where different components can be introduced in the parallel beam by simply moving (sliding) those holders: the exchange is done in a few seconds. The large OHP telescopes being operated with  a Telescope Operator on site, this relaxed a bit the constraints on mechanical and pointing accuracy, as the acquisition and centering into the slit is always checked visually.   

\begin{figure}[ht]
    \centering 
    \includegraphics[width=9.18cm]{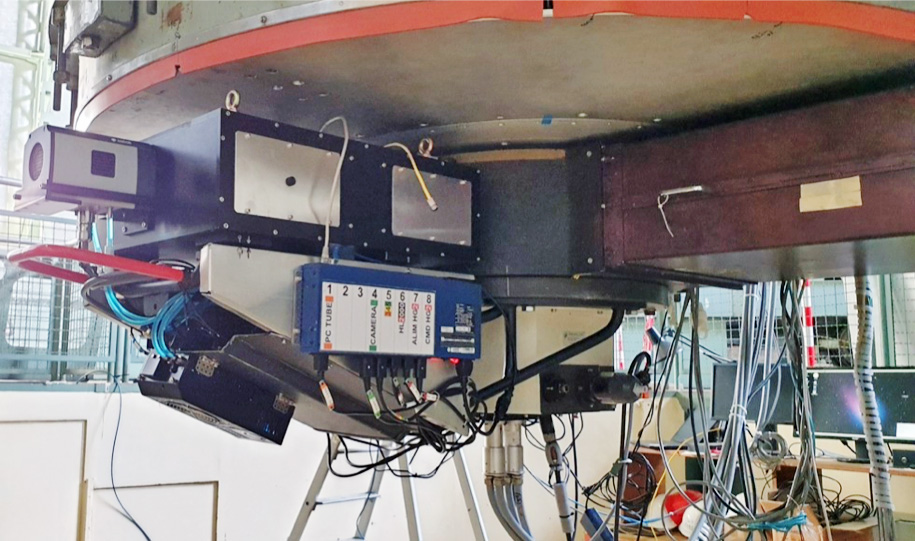}
    \caption[]{MISTRAL (to the left of the image) installed at the folded Cassegrain focus of the 1.93m telescope at OHP}
    \label{fig:FigureMD}
\end{figure}

Regarding the mechanical design, within the T193 Cassegrain Adapter, there is a motorised 45$^{\circ}$ flat mirror which needs to be moved in the beam to
redirect the light onto MISTRAL (at the folded Cassegrain focus) or moved out of the beam  for the other instrument, SOPHIE, which is fed at the direct Cassegrain focus.  
MISTRAL is mounted on one of the side slots of the T193 Adapter, the other one, opposite, being still empty.
Fig.\ref{fig:FigureMD} shows the instrument. The focusing on the slit is done by moving the secondary mirror of the telescope, while the focusing on the detector is done by moving the detector holder. 
A guiding camera is associated directly  to the instrument to minimise mechanical flexures\footnote{\url{http://www.obs-hp.fr/guide/mistral/MISTRAL_spectrograph_camera.shtml#H5}} . 

These modifications lead nevertheless  to a relatively simple instrument, with a long term stability  (see Appendix~\ref{InsStab}) as it is permanently mounted to the telescope and very easy to operate for first time users.

\subsection{Optical Path and Detector}

MISTRAL is mounted at  the folded-Cassegrain focus of the 1.93m telescope via a focal reducer. This focal reducer reduces the beam from  F/15 to F/6 at the entrance slit. At the exit on the detector, the opening is F/3. The entrance slit is presently fixed at a 112 microns width, translating into 1.9 $\arcsec$ 
on the sky, but a variable slit is under construction (the average seeing at OHP is around 2 $\arcsec$).  
At the end, the beam falls on  an ANDOR deep depletion  2K×2K CCD  camera (iKon-L DZ936N BEX2DD CCD-22031) with 13.5 $\mu$ pixels, giving therefore slightly more than 4 pixels per slit width on the detector. This allows for some smoothing in the spectra to increase the signal to noise ratio if needed. The cooling is made by a 5-layer Peltier device. The operating temperature is -90C to -95C. The dark current is lower than 3 electrons/hour/pixel. The sliding plate holding the gratings has presently two grisms, for the blue (roughly 4000-8000~$\AA$) or the red (6000-10000~$\AA$) range, but two more slots are available. The spectral resolution is about 700 at 6000~$\AA$ (see Appendix \ref{VarR} for more details). The FLI filter wheel has 12 positions for 50 mm filters (available presently are : SDSS  g', r', i', z' + Y,
OIII, H$\alpha$, SII, see also appendix \ref{FRs}). Four Thorlabs motors are used to move/remove elements in/out of the optical path: the slit, the grisms, the filters and the calibration mirror. \\
Calibration lights (Hg Ar Xe spectral  lamps and a Tungsten  flatfield lamp) are inserted within the optical path by four optical fibers 
via the calibration mirror which has to be moved in. 

\begin{figure}[ht]
    \centering 
    \includegraphics[width=9.18cm]{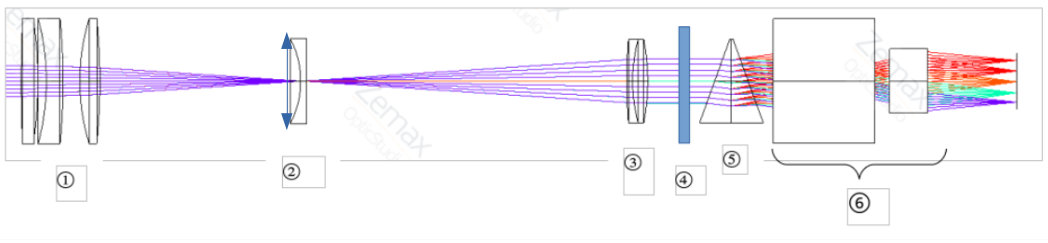}
    \caption[]{MISTRAL optical scheme: (1) focal reducer, (2) field lens -128 mm (with the slit a few mm before), (3)
    achromat collimator f=200mm, (4) filter wheel, (5) VPH 600 tr/mm with two prisms 19,8$^{\circ}$ (blue)/25,6$^{\circ}$ (red), (6) 100mm lens}
    \label{fig:Fig1}
\end{figure}

In the initial design, the Mistral instrument was supposed to be equipped with a wideband camera lens OB V-SWIR F100/2.0  from Optec SPA (https://ir.optec.eu/pdf/C1602.pdf). This camera lens had a coverage from 4000 to 17000~$\AA$ with a good MTF (> 40$\%$ @ 50 lp/mm). Unfortunately, the company was finally unwilling to produce a single lens for us  
and we did not find a reasonable cost alternative on the market. Lacking the 
time and budget to develop a custom lens, we opted to use two different commercial lenses to achieve full coverage of the 4000-10000~$\AA$ spectral band provided by the CCD camera, at the cost of a manual  exchange needed to select the blue or the red setting. 
It is however planned in a second step to replace those two by a single one covering the full range with a good efficiency (see \ref{SingleO}).  
The main characteristics are given in Fig.\ref{fig:Fig1} and Table \ref{tab:Tab1}.

\begin{table*}[ht]
\caption[]{Main characteristics of the MISTRAL instrument \label{tab:Tab1}}
\centering
\begin{tabular}{cc}
\hline
\hline
Wavelength range	&4100-8200~$\AA$ (blue setting) and 5800-9950~$\AA$ (red setting)	\\
Spectral resolution	&R$\sim$700@6000~$\AA$, see also Section \ref{VarR}\\
Fixed Slit Width 	&1.9arcsec	\\
Optical efficiency (Telescope + spectrograph)	&$\sim$0.22@6000~$\AA$ (mean OHP seeing of 2.5arcsec and slit of 2arcsec)	\\
Imaging field of view	&5.1arcmin full light (9arcmin in total)	\\
Filter wheel	&g', r', i', z', Y, H$\alpha$, OIII, SII, H$\beta$, red and blue order separation filters	\\
CCD	& Andor iKon-L 936, 27.6$\times$27.6 mm / 2048$\times$2048 pxs of 13.5$\mu$, Deep Depletion CCD	\\
Spectral Calibration lamps	&Ar/Hg/Xe lamps for wavelength calibration, Tungsten for Flat Fields	\\
Sampling	&0.48arcsec for 13.5 microns pixels	\\
Grism	&Blue: Two prisms at 19.8$^{\circ}$ with VPH at 600 tr/mm @6000~$\AA$ d=50mm	\\
	& Red: Two prisms at 25.6$^{\circ}$ with VPH at 600 tr/mm @9000~$\AA$ d=50mm	\\
Camera lenses	&Blue: Nikon AF-S 100mm F/1.4	\\
	&Red: XENON-EMERALD 2.9/100-L	\\
\hline
\hline
\end{tabular}
\end{table*}

\subsection{Observing modes and available spectral dispersors}
 Several observing modes (Table \ref{tab:Tab3}) are possible with MISTRAL, accessible from a dedicated Graphical User Interface (GUI hereafter), which shows the position of the different elements
in  the optical path. These  are:  the filters (with 12 positions), the spectral dispersors (blue and red VPH gratings, to be used 
 with  the  blue or red camera lens respectively), and the slit (1.9 arc-sec wide). These elements are summarized in Table \ref{tab:Tab1}
and selected according to the different operating modes (imaging, spectroscopy or set-up). 

\begin{figure}[ht]
    \centering 
    \includegraphics[width=9.18cm]{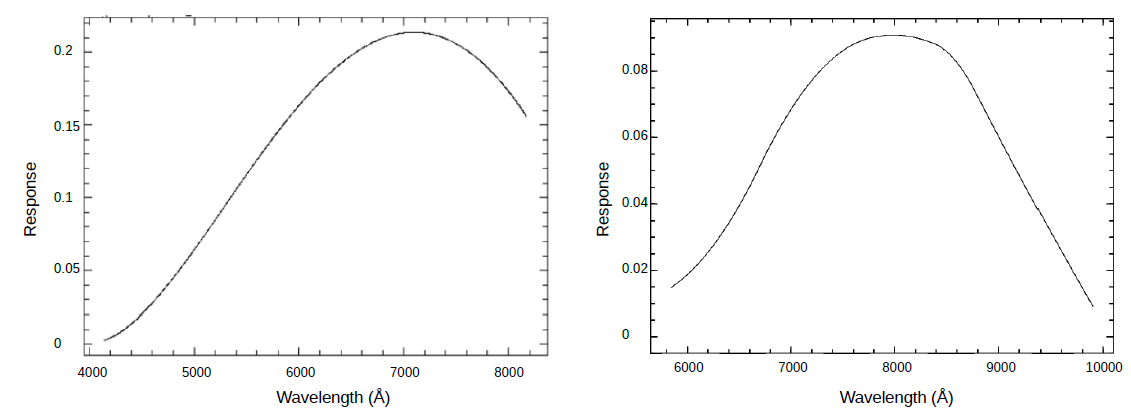}
    \caption[]{Relative spectral response over the CCD 4000-10000~$\AA$ range. Left: blue dispersor / blue lens (from Feige 15
observations). Right: red dispersor / red lens (from BD+26,2606 observations).}
    \label{fig:FigureDRF}
\end{figure}

The dispersor (blue or red) has to be chosen before the beginning of the run in order for the right
camera lens to be mounted during day time. This will change in the future, once a custom-made, wide spectral coverage, unique lens will be available, thus avoiding any dismounting of the instrument. 
In order to measure the CCD spectral response, several standard spectrophotometric stars were
observed. Fig.\ref{fig:FigureDRF} shows the computed spectral response for the blue dispersor / blue lens from
Feige 15 observations \citep{1974ApJ...193..135S}, and for the red dispersor / red lens from BD+26, 2606 observations \citep{1983ApJ...266..713O}. The total efficiency of the system (telescope + instrument + detector) is $\sim$22$\%$  at the peak (6000~$\AA$). 

\begin{table*}[ht]
\caption[]{Main MISTRAL modes: (1): To measure the electronic offset signal.
(2): To center  targets in the image (imaging mode), and to focus the telescope. Fast reading with relatively high
reading noise. These images are not saved.
(3): To take science images.
(4): To take flat field images, to correct the pixel to pixel differential   response. This has to be done
preferentially on the sky (see below), and for each filter, as the  response is very different from one filter to another one.
(5): To locate the slit “x” position on  the CCD before moving the telescope to then center the target within the slit.
The slit is seen thanks to the sky surface brightness. Rapid reading mode.
(6): To move the target at the slit position. Rapid reading mode. Images are not saved.
(7): To take the spectrum of your science target (blue or red disperser). In principle to be done without
other  filters in the beam (but can be done if you  have a special reason to do it), and with an order separation filter for the red disperser.
(8): To take wavelength spectral calibrations in order to transform the  “x” position into wavelength.
(9): To take spectral flat fields  to correct for CCD differential pixel to pixel response. \label{tab:Tab3}}
\centering
\begin{tabular}{cccccccc}
\hline
\hline
Mode                    &  Shutter    &  Reading mode  &    Dispersor&   Slit	&Calib. mirror  &  Wave. calib. lamp  & Spectral flat field lamp	\\
\hline
(1)Bias                 &     Closed         &     50 kHz            &      N/A  &       N/A  &  Off               &    Off                        &   Off	\\
(2)Preview Image        &     Open           &     3 Mhz             &            Off  &       Off  &  Off               &    Off                        &   Off	\\
(3)Science Image        &     Open           &     50 kHz            &            Off  &       Off  &  Off               &    Off                        &   Off	\\
(4)Imaging Flats        &     Open           &     50 kHz            &            Off  &       Off  &  Off               &    Off                        &   Off	\\
(5)Search Slit Position &     Open           &     3 Mhz             &            Off  &       On   &  Off               &    Off                        &   Off	\\
(6)Centering on the slit&     Open           &     3 Mhz             &            Off  &       Off  &  Off               &    Off                        &   Off	\\
(7)Science Spectrum     &     Open           &     50 kHz            &            On   &       On   &  Off               &    Off                        &   Off	\\
(8)Calibration Arc      &     Open           &     50 kHz            &           On   &       On   &  On                &    On                         &   Off	\\
(9)Calibration Tungsten &     Open           &     50 kHz            &            On   &       On   &  On                &    Off                        &   On	\\
\hline
\end{tabular}
\end{table*}

\subsection{Exposure Time Calculators}

Exposure time calculators are available remotely. One for spectroscopy: ETC1\footnote{\url{http://www.obs-hp.fr/guide/mistral/MISTRAL_ETC1.shtml}} and one for imaging: ETC2 \footnote{\url{http://www.obs-hp.fr/guide/mistral/MISTRAL_ETC2.shtml}}. The details of the needed parameters to enter into the calculators are described in the Appendix \ref{ETCs}, and some results are displayed there also.  Here, we only give a quick flavor of what can be achieved. \\
For spectroscopy,  we summarize in Table \ref{tab:Tab4} the V band magnitudes detected with a total exposure time
of 1 hour, with a S/N of 3, for point-like objects, under a median seeing (for OHP) of 2.5 arcsec,
and with moon conditions: 135$^{\circ}$ from moon, 25$\%$ illumination.

\begin{table*}[ht]
\caption[]{Typical spectroscopic limiting magnitudes corresponding to a total exposure time
of 1 hour, with a S/N of 3, for point-like objects, under a median seeing of 2.5 arcsec,
and with moon conditions: 135$^{\circ}$ from moon, 25$\%$ illumination. Values are given for pure absorption line spectra (blue dispersor),
pure absorption line spectra (red dispersor), emission line spectra (blue dispersor), and emission line line spectra (red dispersor) \label{tab:Tab4}}
\centering
\begin{tabular}{ccccc}
\hline
\hline
             &  Abs. lines Blue dispersor &  Abs. lines Red dispersor  &   Em. Lines Blue dispersor  &    Em. Lines Red dispersor \\
\hline
V magnitude  &  $\leq$19.5         &  $\leq$19.5                   &   $\leq$20.5                     &    $\leq$20			     \\
\hline
\end{tabular}
\end{table*}

For imaging,  ETC2 gives the exposure time needed to detect objects at a given magnitude with the requested S/N. While the details of the procedure are fully described in the Appendix, we summarize the important results in Table \ref{tab:Tab5}. 

\begin{table}[ht]
\caption[]{90$\%$ completeness for point sources under average imaging observing conditions at OHP. \label{tab:Tab5}}
\centering
\begin{tabular}{cc}
\hline
\hline
g'                &   20min: g'$\sim$20.5 \\
r'                &  10min: r'$\sim$20. \\
i'                &  10min: i'$\sim$21. \\
z'                &  10min: z'$\sim$19.5 \\
Y                 &  10min: Y$\_{AB}$$\sim$18.5 \\
\hline
\end{tabular}
\end{table}

See also \url{http://www.obs-hp.fr/guide/mistral/CookBook_main.pdf} (page 23) for the minimal possible exposure times.

\subsection{Data Archival}
In addition to their local storage in the MISTRAL computers at the telescope, raw MISTRAL data are automatically archived  within a database\footnote{\url{https://data.lam.fr/mistral/home}} hosted by the CeSAM\footnote{\url{https://www.lam.fr/servive/cesam/preentation/}}. Raw data are visible but not accessible during a proprietary period of 12 months to people other than principal
investigators. All calibration data are immediately public. A possibility is also offered to the observers to store/make available their final reduced data and added
values through the GASPIC national service\footnote{\url{https://gaspic.osupytheas.fr}}.

\subsection{Expected optical improvement}
\label{SingleO}

In order to increase the instrument's efficiency, and avoid change-overs of optics when changing from blue to red settings (or vice-versa),  we are  preparing  a single camera lens covering
 the full MISTRAL spectral range: it is expected to be installed on the instrument in 2024/2025.

The new, custom-made camera design consists of 5 lenses with 2 aspheric surfaces and uses low-dispersion OHARA S-FPL53 glass for the strong positive lenses. It is optimized to operate in the full working spectral range with a dispersed beam, which implies a forward-shifted entrance pupil.

The expected performances of the lens in development  are compared to those of the commercial lenses which are currently used in the instrument. In the actual blue-visible setting, an FX AF-S NIKKOR 105mm f/1.4 lens is used  and its module transfer function (MTF hereafter) is available from the manufacturer at two reference spatial frequencies. In order to make a correct comparison, the calculations in the visible were made for the \textit{F}, \textit{d} and \textit{C} wavelengths (4860-6560~$\AA$) and the field edge corresponds to $11^{\circ}$ in each case. As the throughput curve for this Nikon lens is not available, it was estimated from the known number of surfaces and groups in the design, presuming that the anti-reflective (AR) coating has a residual reflectance of $1\%$ (similar to that in the custom design, see also \url{http://www.obs-hp.fr/guide/mistral/CookBook_main.pdf} (page 20) and the glass types composition is close to that used in [U.S.Patent 5572277]. The commercial lens data are taken from the www\footnote{\url{https://photographylife.com/lenses/nikon-af-s-nikkor-105mm-f1-4e-ed/3}}. It comes out that the throughput increases 
from about  75$\%$  to about 95$\%$ at 5000, 6000 or 7000~$\AA$ and markedly from 19$\%$ to 77$\%$ at 4000~$\AA$, while the MTF is only marginally decreasing.

In the same way the comparison was done for the Schneider Emerald 2.9/100 F-LD lens, which is actually used in the red range. This time both the MTF and throughput are known from the product's  datasheet\footnote{\url{https://schneiderkreuznach.com/application/files/3316/4725/4957/EMERALD_29_100_F-LD_1070506_datasheet.pdf}}. 
Here the foreseen transmission  is similar with the present one at 7000~$\AA$ ($93\%$), and slightly lower at 8000~$\AA$ (90$\%$  instead of $95\%$), with a few percent decrease of the MTF at 40l/mm. 
However, it should also be noticed that the aperture of the current Schneider lens is insufficient and results in significant vignetting towards the edges of the spectrum, with a maximum loss of 40$\%$ at 6000~$\AA$ and 10000~$\AA$. Thus,  this vignetting loss will disappear with the new camera lens.

  Note  that the nominal f-number, design wavebands and the pupil positions are different for all these lenses, so the comparison is not straightforward. However, it indicates that the custom design is notable for an increased throughput in the shortwave visible range and the contrast at mid-spatial frequencies remains high even for a faster beam. 

% \begin{table*}
%    \centering
%    \caption{Comparison of throughput and resolution  between old and new   % lens design\label{tab:TabEM}}
%    \label{tab:Optical_performance}
%    \begin{tabular}{|c|c|c|c|}
%        \hline
%         &  Custom camera lens &  \thead{FX AF-S NIKKOR\\ 105mm f/1.4}  &   % \thead{Schneider Emerald\\ 2.9/100 F-LD}    \\
%          \hline
%        Wavelength, nm &  \multicolumn{3}{|c|}{Throughput} \\
%         \hline
%         $400$ &  $77.4\%$ &  $18.6\%$ & $52\%$ \\
%         $500$ &  $93.1\%$ &  $72.5\%$ & $92\%$ \\
%         $600$ &  $94.1\%$ &  $74.7\%$ & $93\%$ \\
%         $700$ &  $93.8\%$ &  $75.2\%$ & $93\%$ \\
%         $800$ &  $90.3\%$ &  n/a      & $95\%$ \\
%         \hline
%         Spatial frequency, 1/mm& \multicolumn{3}{|c|}{Contrast in field   % centre (edge)} \\
%         \hline
%         $30$ & $0.912(0.59)$  & $0.96 (0.63)$  & - \\
%         $40$ & $0.575(0.469)$ & -              & $0.63 (0.57)$ \\
%         \hline
%    \end{tabular}
%
% \end{table*}

\section{On-sky validations}
In order to give to the reader a real-life idea of what is possible with MISTRAL, we report now 
salient (unpublished) examples of
science done with MISTRAL observations during its first two years of exploitation.
\label{OSV}
Following the main science goals of the instrument, as described in the introduction, various types of objects were observed during several test periods, starting in April 2021 and lasting until around end of 2022, with progressively more science taking over the tests themselves. We describe below some of the results obtained so far. 

We give in Table \ref{tab:summary} the observing conditions of the spectra shown in Figures \ref{fig:NovaCas}, \ref{figureBS}, \ref{fig:LkHA_324SE}, and \ref{fig:FigureC2022E33}.

\begin{table*}[!]
    \centering
     \caption{Observing details for the targets described in section \ref{OSV}. Moon illumination is given in percentage along with the distance to the target in degrees.}
   \begin{tabular}{|c|c|c|c|c|c|c|}
    \hline
        Target Name & Observation Date & Seeing (arcsec) & Moon & Exposure Time & Air Mass & Dispersor  \\
        \hline\hline
         NovaCass 2021 & 29/06/2022 & 3.0 & 1$\%$@80$^{\circ}$ & 2$\times$600sec & 1.2 & red \\\hline
         NovaCass 2021 & 01/07/2022 & 5.1 & 4$\%$@86$^{\circ}$ & 1$\times$800sec & 1.1 & blue \\\hline
          LkH$\alpha$~324SE & 08/12/2021 & 3.2 & 26$\%$@72$^{\circ}$ & 2$\times$900sec & 1.2 & red\\\hline
         2G1045666+0128085 & 02/07/2022 & 2.2 & 8$\%$@98$^{\circ}$ & 1$\times$900sec & 1.05 & blue \\\hline
         C/2022 E3 (ZTF) & 10/02/2023 & 1.9 & 78$\%$@123$^{\circ}$ & 10$\times$240sec + 2$\times$120sec & 1.1 & blue \\\hline
    \end{tabular}
    \label{tab:summary}
\end{table*}

\subsection{Emission lines stars: example of NovaCas 2021}

\begin{figure*}
    \centering
    \includegraphics[width=18cm]{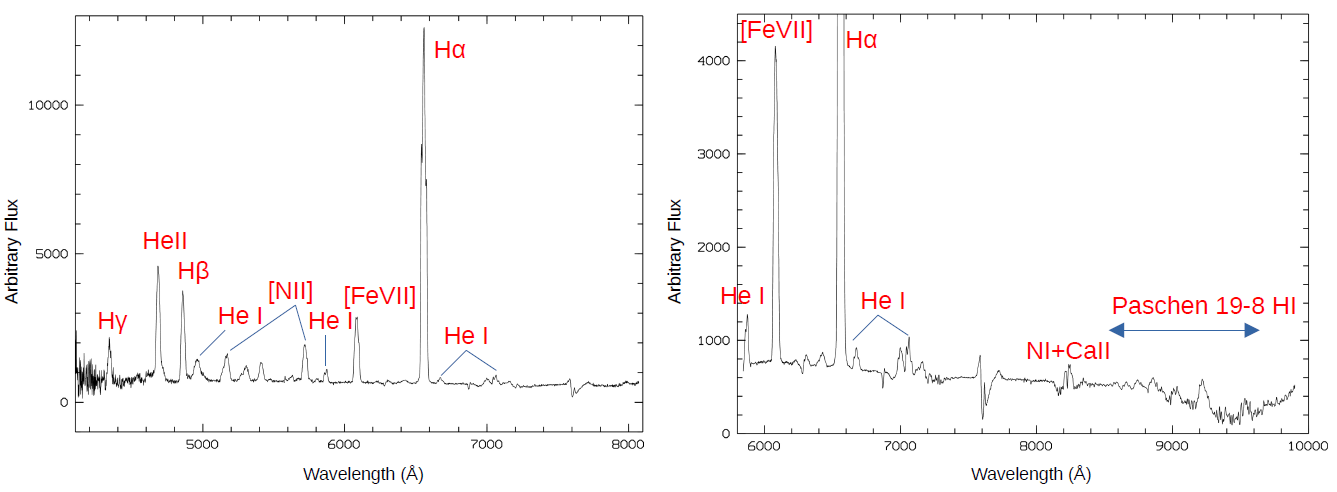}
    \caption{.
       Spectra of V1405Cas = NovaCas2021, obtained in June 2022, with the blue setting (left) and the red setting (right, enhanced y scale). } 
    \label{fig:NovaCas}
\end{figure*}

V1405 Cas was discovered on March, 18th 2021 and could be observed during the first commissioning nights of the instrument, in April and July 2021. 
For instance, the spectra obtained on July 5th, 2021  (that is about  100 days after maximum),  still showed  PCygni profiles in the H and He Balmer lines, sign  of the expending envelope, but also some NII lines and some FeII lines, indicating that the object was transiting from the He/N phase to the FeII phase.   In 
Figure \ref{fig:NovaCas} we show two spectra of this slow Nova,  observed  later with both the blue and red MISTRAL grisms, on June30th, and June 28th, 2022, respectively, that is about 415 days after maximum. The PCygni  profiles do not appear anymore in those spectra,  and the line profiles do not show yet a rectangular shape either, confirming the fact that we deal  with an FeII type. 
The linewidths of the strong lines are measured at 1800 km/s FWHM. 
Besides the lines of H and He, most conspicuous are the strong lines of [FeVII], in particular those at 6086 and 5721~$\AA$. On the other hand, the absence of the usually strong OI lines 8446~$\AA$ (and  the less strong 7773~$\AA$) is surprising: we have to wait that the Nova has reached the full nebular phase to make a proper abundance analysis.  

\subsection{General stellar classification: bow shock stars}

\begin{figure}[ht]
    \centering \includegraphics[width=6.5cm,angle=270]{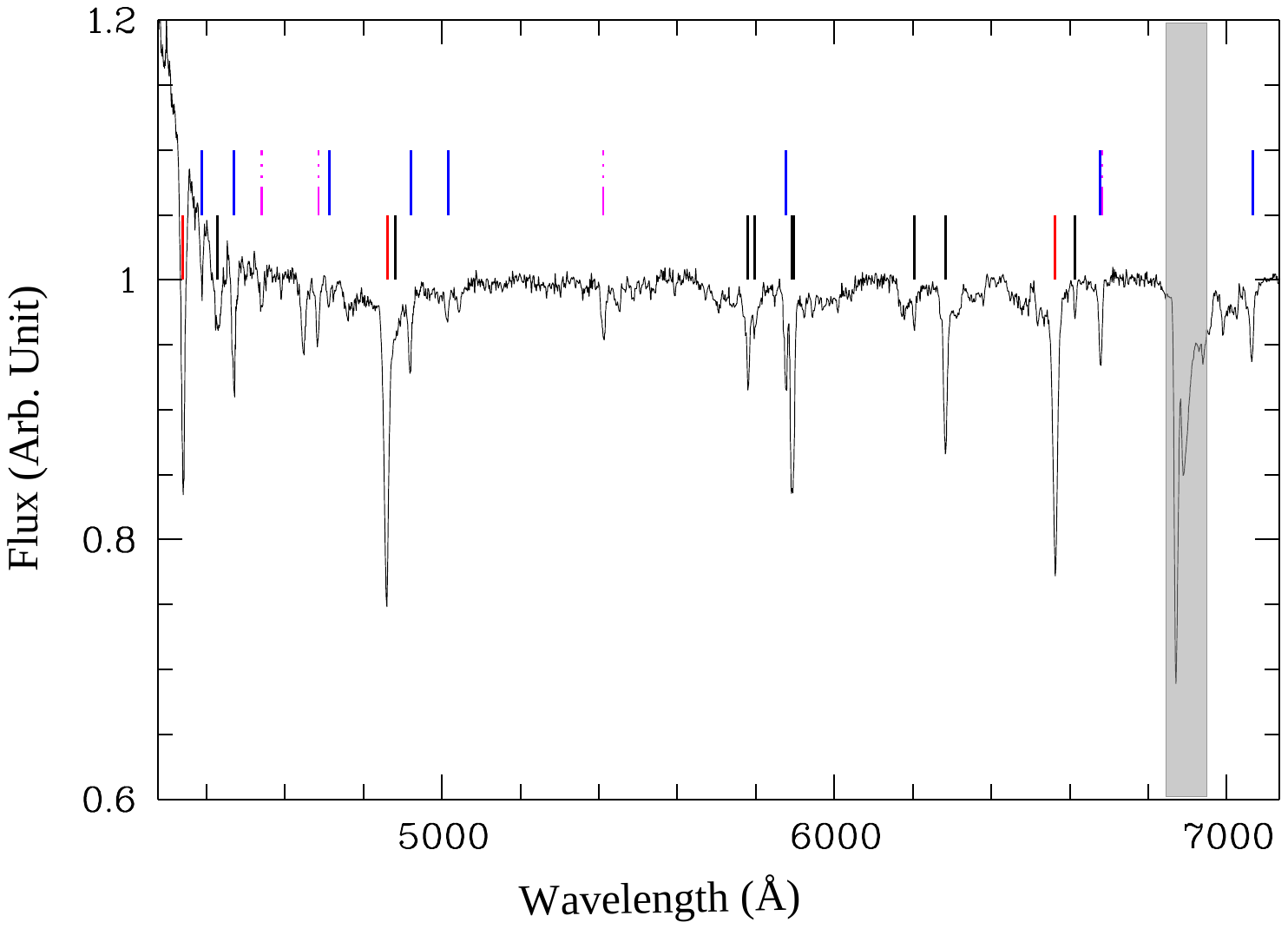}
    \caption[]{Continuum normalised spectrum observed with MISTRAL of the candidate BS star 2G1045666+0128085. 
The position of the main lines used for the classification (magenta, blue, and red dashes correspond to HeII, HeI and Balmer lines respectively) and the diffuse interstellar bands (blacks dashes) are indicated. The grey vertical band underlines the night-sky feature. }
    \label{figureBS}
\end{figure}

Bow-shocks (BS) are arc-shaped structures located ahead of a star and generally observed at mid-to-far IR
wavelength. They are expected to result from the interaction of the stellar wind with
the ambient interstellar medium (ISM) when the star has a relative supersonic motion respectively to the ambient
medium. Because the bow-shock driving star is expected to be a O or B type star we led a  spectroscopic follow-up of a sample of 47 bow-shock stellar candidates
selected from the \cite{Jayasinghe19} catalog and selected to be observable with the MISTRAL instrument. Figure \ref{figureBS} shows a typical spectrum for a V=9.7 star.

The spectral classification of stars is commonly done from lines in the 4000-5000~\AA~domain but, due to the lower sensitivity of the instrument 
in this spectral range, we performed the spectral classification from lines in the red part (4500-7000~\AA) of the spectrum (Mk standards in the red have been provided by, e.g. \cite{1994PASP..106..382D}).
We find that among the 47 candidates we have: 2 unclassifiable stars, 3 cool stars, 1 A-type star, 10 O stars and 31 B (mainly giants
and super-giants) stars which allows us to confirm that bow-shocks are mainly driven by hot stars. The details of these results, complemented with transverse velocity study (based on Gaia-DR3 astrometric data) of the stars, can be found in Russeil et al. (in preparation).

\subsection{The red domain towards 1$\mu$: the variable, pre-main sequence star LkH$\alpha$~324SE in Lynds~988}

\begin{figure}[!t]
\centering
\includegraphics[width=1\columnwidth]{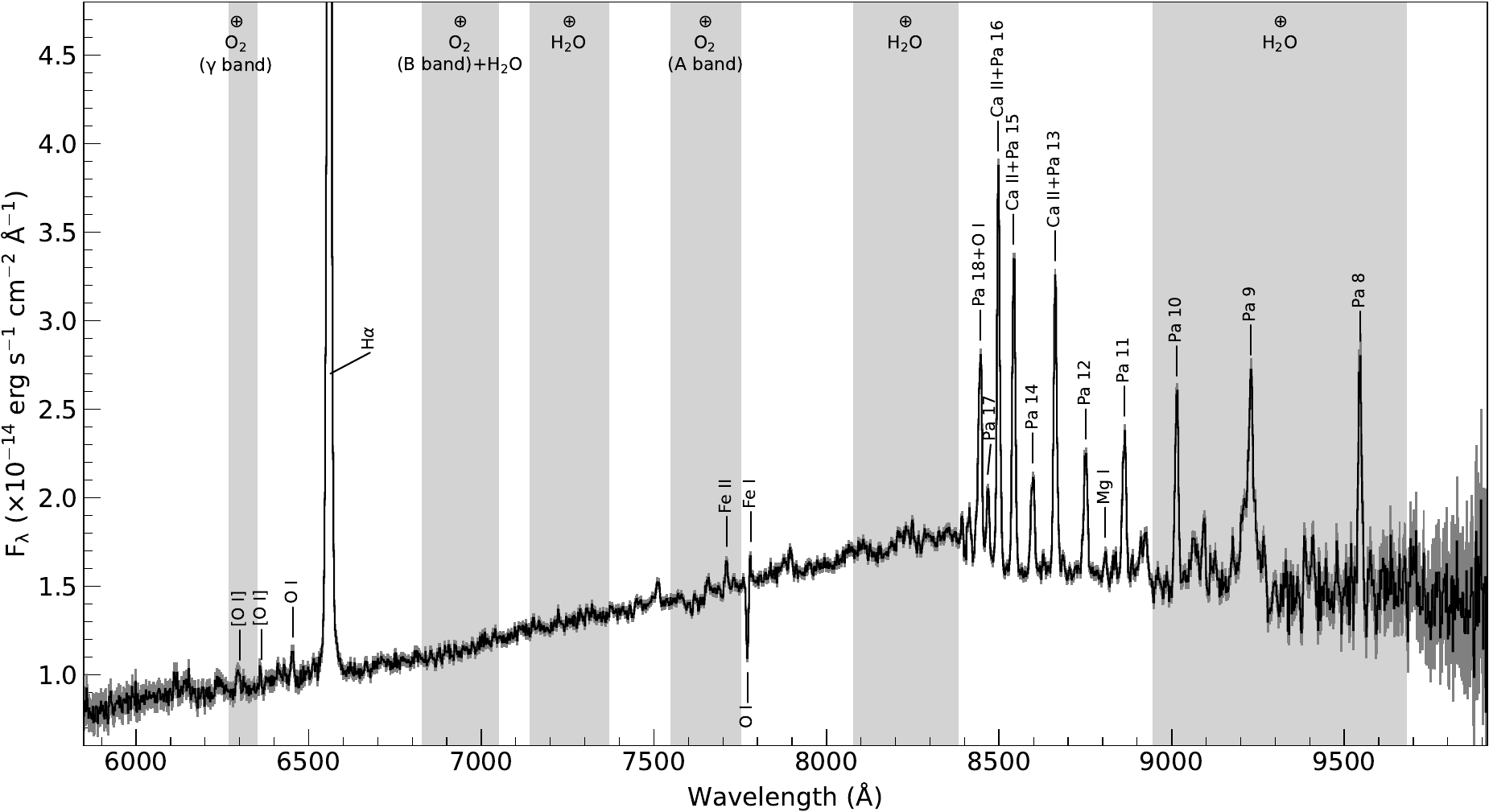}
\caption{MISTRAL spectrum of LkH$\alpha$~324SE with telluric corrections, 
obtained with the red configuration 
and a single exposure of 900~s (among the two observed)
on the night of 2021 Dec.~8. 
The flux error-bars are plotted in gray.
The vertical light-gray stripes indicate the atmospheric 
absorption bands of water and molecular oxygen.}
\label{fig:LkHA_324SE}
\end{figure}

Gaia21fji\footnote{\url{http://gsaweb.ast.cam.ac.uk/alerts/alert/Gaia21fji/}}
(a.k.a. AT2021aftk\footnote{\url{https://www.wis-tns.org/object/2021aftk/discovery-cert}}
in the IAU designation) was reported on 2021 Nov.~6 
as the "brightening of a variable red Gaia source, candidate YSO"
\citep{hodgkin21}.
We obtained a classification spectrum on 2021 Dec.~8 with MISTRAL 
in the red configuration \citep{adami21}.
In fact, Gaia21fji corresponds 
to the emission-line star LkH$\alpha$~324SE 
(\citealt{herbig72,chavarria-k83};  
a.k.a.\ HBC~727 in \citealt{herbig88}), 
which is a member of the pre-main sequence 
population of the dark cloud Lynds~988 
located at $\sim$600~pc \citep{herbig06},
therefore, it is not a new candidate YSO identified 
by Gaia. The spectrophotometric standard Feige~15 
was observed after LkH$\alpha$~324SE
(Appendix~\ref{appendix:telluric}).

The spectrum displays many strong emission lines 
above a red continuum (Fig.~\ref{fig:LkHA_324SE}): 
H$\alpha$, Paschen (Pa) lines, 
the Ca\,{\sc ii} infrared triplet (IRT), 
and the O\,{\sc i} $\lambda$8446. 
Fainter emission lines of Mg\,{\sc i}, 
Fe\,{\sc i}, and Fe\,{\sc ii} are also detected, 
plus forbidden O\,{\sc i} lines.
The O\,{\sc i} $\lambda$7772 absorption triplet 
is detected. The Ca\,{\sc ii}~IRT in emission is an indicator 
of chromospheric activity and potentially of accretion 
\citep{mohanty05, yamashita20}.

\begin{figure}[!t]
\centering
\includegraphics[width=\columnwidth]{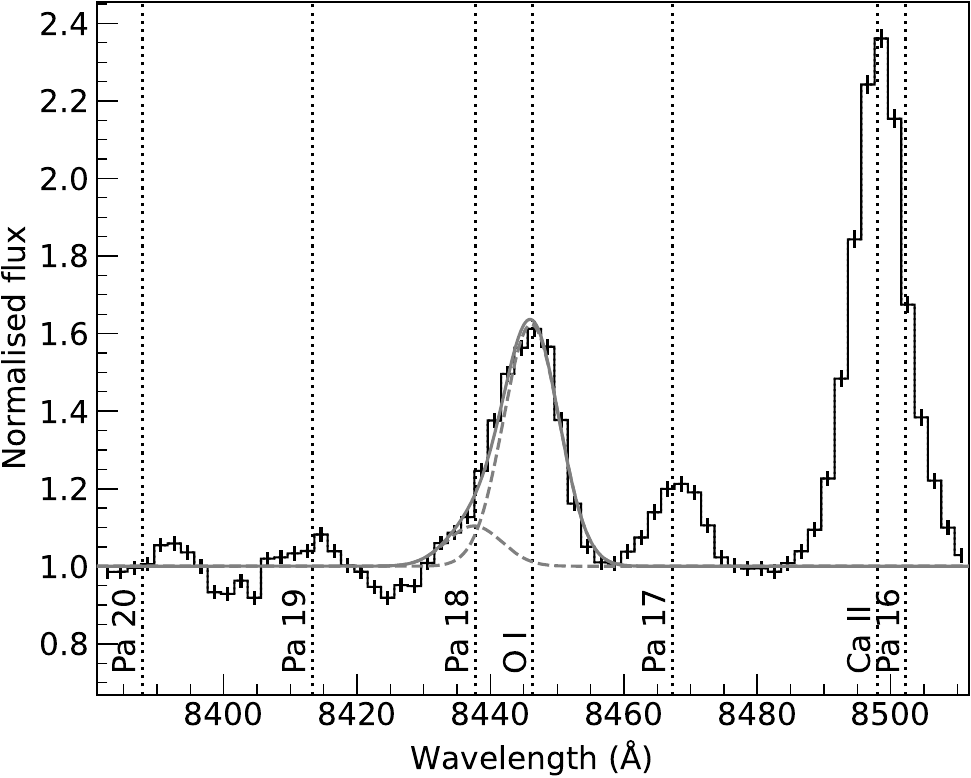}
\caption{Pa 18 $\lambda$8437.75 and 
O\,{\sc i} $\lambda$8446.76 from LkH$\alpha$~324SE.
The grey solid line is the combined Gaussian best-fit 
of these two lines.
The grey dashed lines are the individual Gaussian best-fits 
of these two lines, confirming 
the detection of the Pa~18 line (see text for details).}
\label{fig:Pa18_OI}
\end{figure}

The emission lines Pa 8--12, 14 and 17 are detected.
Pa~8--10 are located in the MISTRAL spectrum region that is 
strongly affected by the fringing (see also \url{http://www.obs-hp.fr/guide/mistral/CookBook_main.pdf} page 16 and \url{http://www.obs-hp.fr/guide/mistral/Test_report.pdf} page 15) and the water absorption-band, 
therefore, the measured equivalent widths (EW hereafter) could be sensitive to the determination 
of the pseudo-continuum and the correction of the telluric 
absorption. 
Pa~16 $\lambda$8502.27, 
Pa~15 $\lambda$8545.17, 
and
Pa~13 $\lambda$8664.80
cannot be resolved with MISTRAL from  
the emission lines  
$\lambda$$\lambda$8498.02, 8542.09, 8662.14, 
respectively, 
of the Ca\,{\sc ii}\xspace IRT
(see bottom panel of Fig.~\ref{appendix_fig:R}).
However, Pa 18 $\lambda$8437.75 and 
O\,{\sc i} $\lambda$8446.76 
are 9.01~\AA\ apart, 
which is just large enough to be barely resolved 
as MISTRAL FWHM is 8.5~\AA\ at this wavelength 
(see bottom panel of Fig.~\ref{appendix_fig:R}).
Fig.~\ref{fig:Pa18_OI} shows the MISTRAL spectrum 
centered on the O\,{\sc i} $\lambda$8446.76 
with a hint of Pa~18 on its red wing. 

\begin{figure}[!t]
\centering
\begin{tabular}{cc}
\includegraphics[width=\columnwidth]{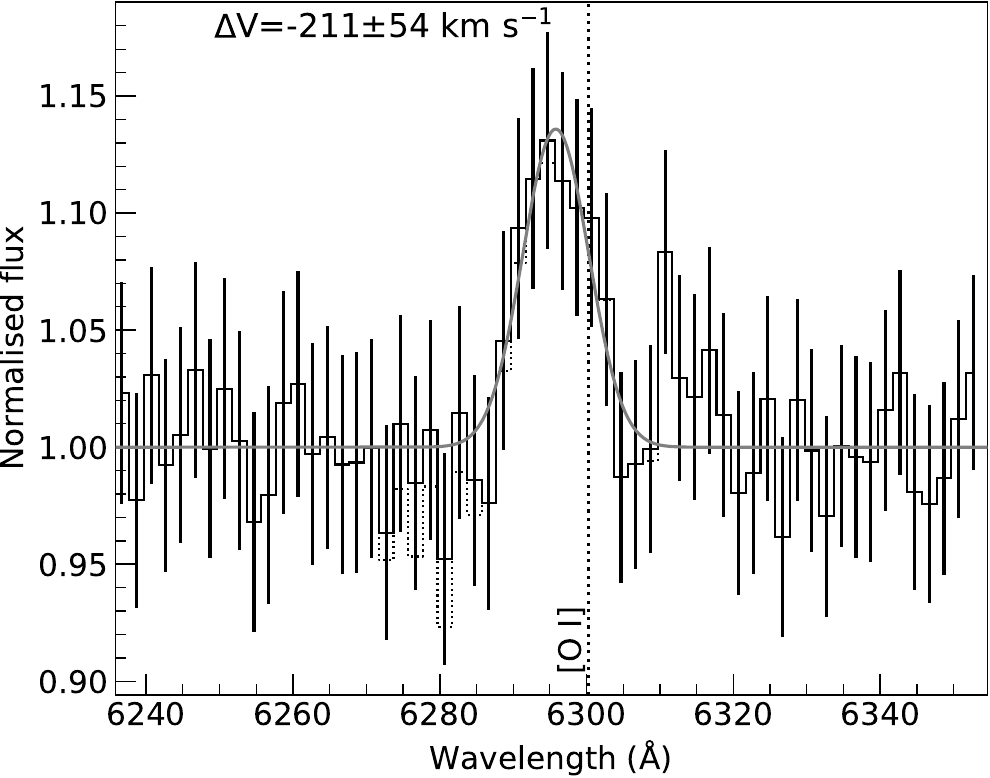}\\
\includegraphics[width=\columnwidth]{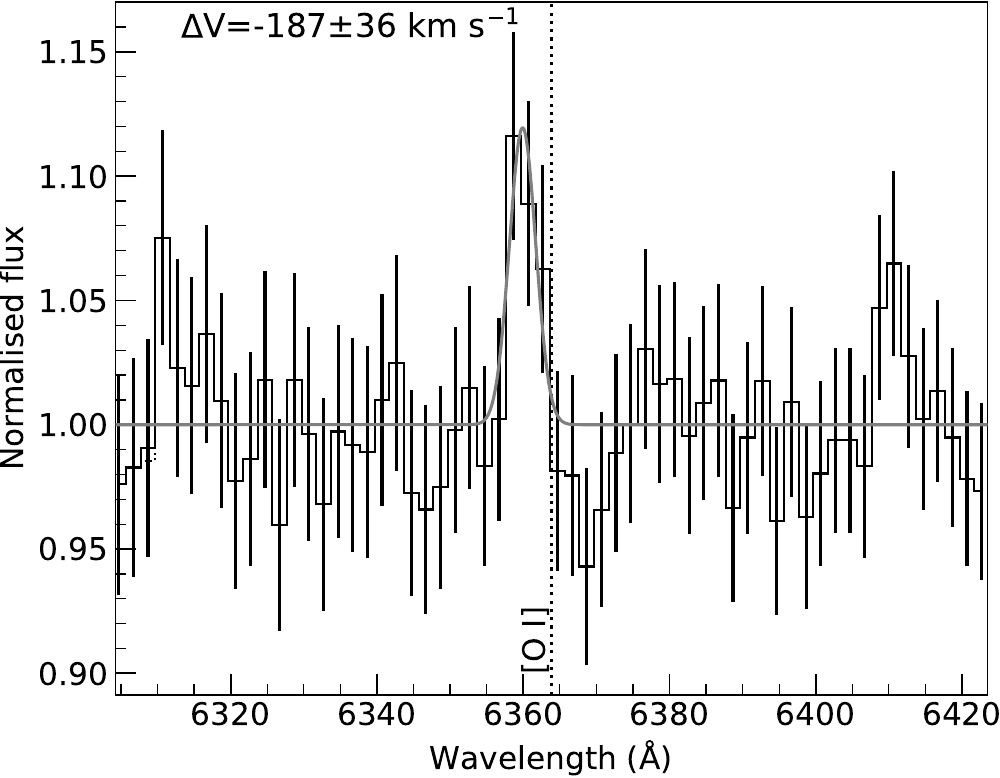}\\
\end{tabular}
\caption{Faint, blueshifted [O\,{\sc i}] emission lines from LkH$\alpha$~324SE.
In the top panel, the dotted step line is the 
normalised spectrum with no correction 
of the atmospheric O$_2$ $\gamma$ band. 
The gray line is the Gaussian fit. 
The quoted error of the velocity shift includes 
quadratically the wavelength calibration error.}
\label{fig:[OI]}
\end{figure}

\cite{herbig06} have also identified 
the [S\,{\sc ii}], [O\,{\sc i}], and [N\,{\sc ii}] lines
from LkH$\alpha$~324SE;
the [O\,{\sc i}] $\lambda\lambda$6300, 6363 lines have the same shape 
than the [S\,{\sc ii}] $\lambda\lambda$6717, 6730 lines, 
which display a narrow component at $-18$~km~s$^{-1}$ superposed
to a very broad asymmetric line peaking near $-200$~km~s$^{-1}$
(their Fig.~12, bottom panel).
The [O\,{\sc i}] high- and low-velocity components of T~Tauri stars 
are associated to micro-jet and disk wind, 
respectively \citep{hartigan95}.
The [N\,{\sc ii}] $\lambda\lambda$6548, 6583 lines cannot be resolved 
with MISTRAL from the broad wings of the H$\alpha$ line.
The [S\,{\sc ii}] $\lambda\lambda$6717, 6730 lines are not detected with MISTRAL, 
but our $1.9\arcsec$-slit was oriented NS, 
not at position angle of 309$^\circ$ along the NW flow as in \cite{herbig06},
which may dilute the observed flux 
from the base of the NW flow.
Fig.~\ref{fig:[OI]} shows the regions of the [O\,{\sc i}] 
$\lambda$$\lambda$6300.23, 6363.88 lines with two faint 
emission lines with EW of $-1.6$ and $-0.3$~\AA, respectively.
These faint lines are blueshifted, with velocity shifts that 
are consistent with the high-velocity, broader component 
of these lines. 
Therefore, MISTRAL can also detect faint forbidden [O\,{\sc i}] lines 
from this pre-main sequence star.

\subsection{Test of the MISTRAL spectral resolution around the H$\alpha$ line in LkH$\alpha$~324SE}
% Ha fit

\begin{figure}[!h]
\centering
\includegraphics[width=\columnwidth]{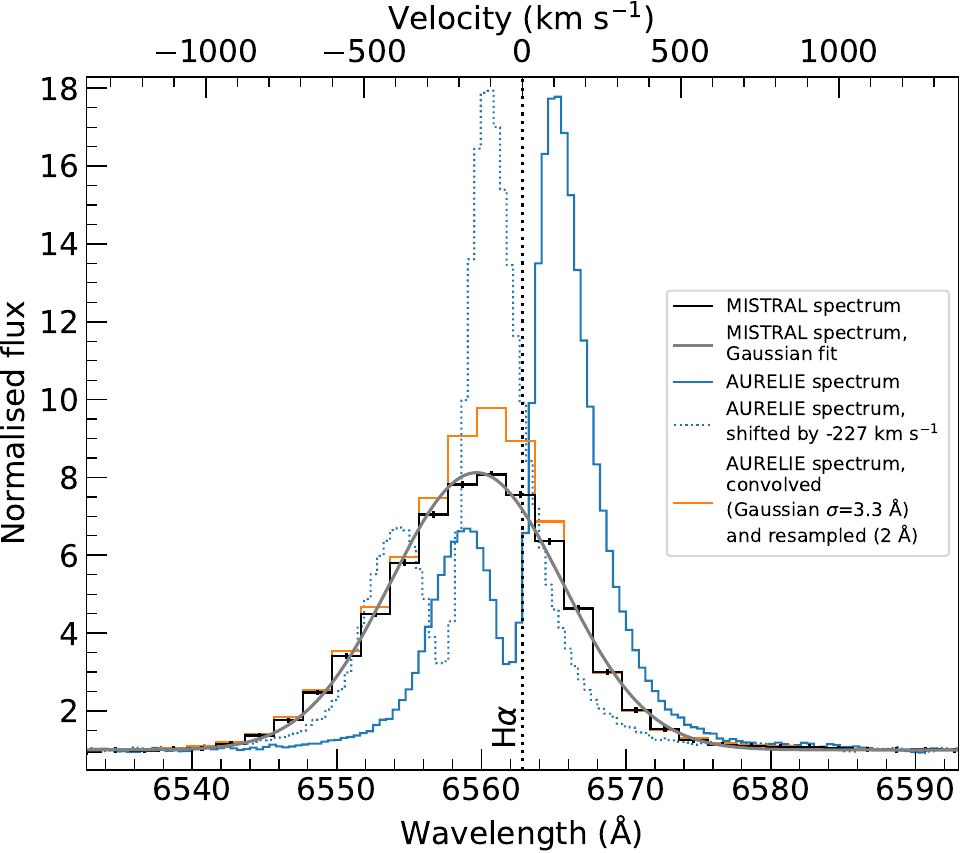}
\caption{Broad and variable, H$\alpha$ emission line from LkH$\alpha$~324SE. 
The black data with error bars are the 
normalised MISTRAL spectrum.
The grey line is the Gaussian fit of the emission line. 
The solid blue line is the medium-resolution ($R$$\sim$5,000) spectrum 
obtained, 12 days after the MISTRAL observations, 
with the OHP 1.52m telescope with the AURELIE spectrograph.
The dotted blue line is the same spectrum
shifted by $-227$~km~s$^{-1}$ to match the 
position of the emission line observed with MISTRAL.
The orange line is the convolved (Gaussian with $\sigma=3.3$~\AA) 
and resampled (2~\AA\ spectral~bins) AURELIE spectrum 
to match MISTRAL spectral resolution and sampling.}
\label{fig:Halpha}
\end{figure}

% 
In  MISTRAL's observations of LkH$\alpha$~324SE, the H$\alpha$ line is strong 
with an EW of -107~\AA,
which is well above the maximum H$\alpha$ EW of -20~\AA\ 
that can be produced by active stellar chromosphere 
in later M spectral-type, therefore, the origin of this 
emission is from accretion shock at the stellar surface 
\citep{muzerolle01}.
Moreover, the H$\alpha$ line is broad 
with FWHM(H$\alpha$)=$13.98\pm0.05$~\AA\
(Fig.~\ref{fig:Halpha}), 
which is well larger than 
 MISTRAL's spectral resolution of 7.9~\AA\ 
for the red configuration (Appendix Eq.~\ref{equation:Red}).
The (deconvolved) full width at 10\% of the H$\alpha$ peak is 
$961\pm5$~km~s$^{-1}$ (Appendix Eq.~\ref{equation:velocity}), 
which is larger than the minimum value of 270~km~s$^{-1}$ 
used as accretion criteria in low-mass stars \citep{white03}. 
We note small residuals between the MISTRAL H$\alpha$ line 
and the Gaussian fit
that suggest for this line an asymetric profile. 
Indeed, the high-resolution ($R$$\sim$45,000) spectra of 
LkH$\alpha$~324SE,
obtained 
on 2003 Jul.~6 and Dec.~13 
with the HIRES spectrograph 
at the Keck I telescope on Mauna Kea, 
shows the H$\alpha$ line 
(Fig.~10, top panel of \citealt{herbig06})
with an EW of $-170$~\AA,
wings extending to at least $\pm$550~km~s$^{-1}$, 
and an absorption component
at $-114$~km~s$^{-1}$, 
usually
associated with a stellar wind, 
between the secondary and main line-peaks 
at -200 and 2~km~s$^{-1}$, respectively.
We note that the secondary peak is less than half 
the strength of the primary peak, classifying this 
H$\alpha$ line as a type~IIIB profile, which is the 
most common (33\%) in T~Tauri stars, whereas 
it is three times less common (11\%)
in Herbig Ae/Be stars 
\citep{reipurth96}.

A medium-resolution ($R$=5,000) spectrum 
in the blue 
of LkH$\alpha$~324SE
%wavelength range 4710--6750~\AA\
was obtained by C.A.\ 12~days after the MISTRAL observation 
on the night of 2021 Dec.~20, with the AURELIE spectrograph 
at the OHP 1.52m telescope \citep{gillet94}.  
This AURELIE spectrum will be reported elsewhere. Here, 
we focus only on the H$\alpha$ line observed 
with AURELIE (Fig.~\ref{fig:Halpha}).
This type~IIIB-profile line displays: 
an EW of $-119$~\AA, i.e.,
43\% fainter and 10\% stronger than during the 
HIRES and MISTRAL, respectively, observations; 
the absorption 
component
at $-52$~km~s$^{-1}$, 
i.e., twice-less blueshifted 
than observed 18 years before; 
the secondary and main line-peaks at -174 and 112~km~s$^{-1}$,
respectively, 
i.e., less blueshifted and much more redshifted,
respectively,
than observed 18 years before.
The $I_\mathrm{max}/40$ velocities \citep{reipurth96} for the 
blue and red wings are -507 and 601~km~s$^{-1}$, respectively.
We use the shape of the AURELIE H$\alpha$ line as template 
for comparison with the MISTRAL H$\alpha$ line. 
After convoluting (Gaussian with $\sigma=3.3$~\AA\ from quadratic combination of resolution) 
and resampling (2~\AA\ spectral-bins) this template,
the best-fit match with the MISTRAL H$\alpha$ line is obtained 
for a velocity shift of about $-227$~km~s$^{-1}$.
However, this large value of the required velocity shift suggests 
on the timescale of 12 days
a strong variation of the shape of the H$\alpha$ line, 
including for instance, a more blueshifted 
absorption component
during the MISTRAL observation.

% Paschen lines

\subsection{Solar System objects: the C/2022 E3 (ZTF) comet}

\begin{figure}[ht]
    \centering 
    \includegraphics[width=9.cm]{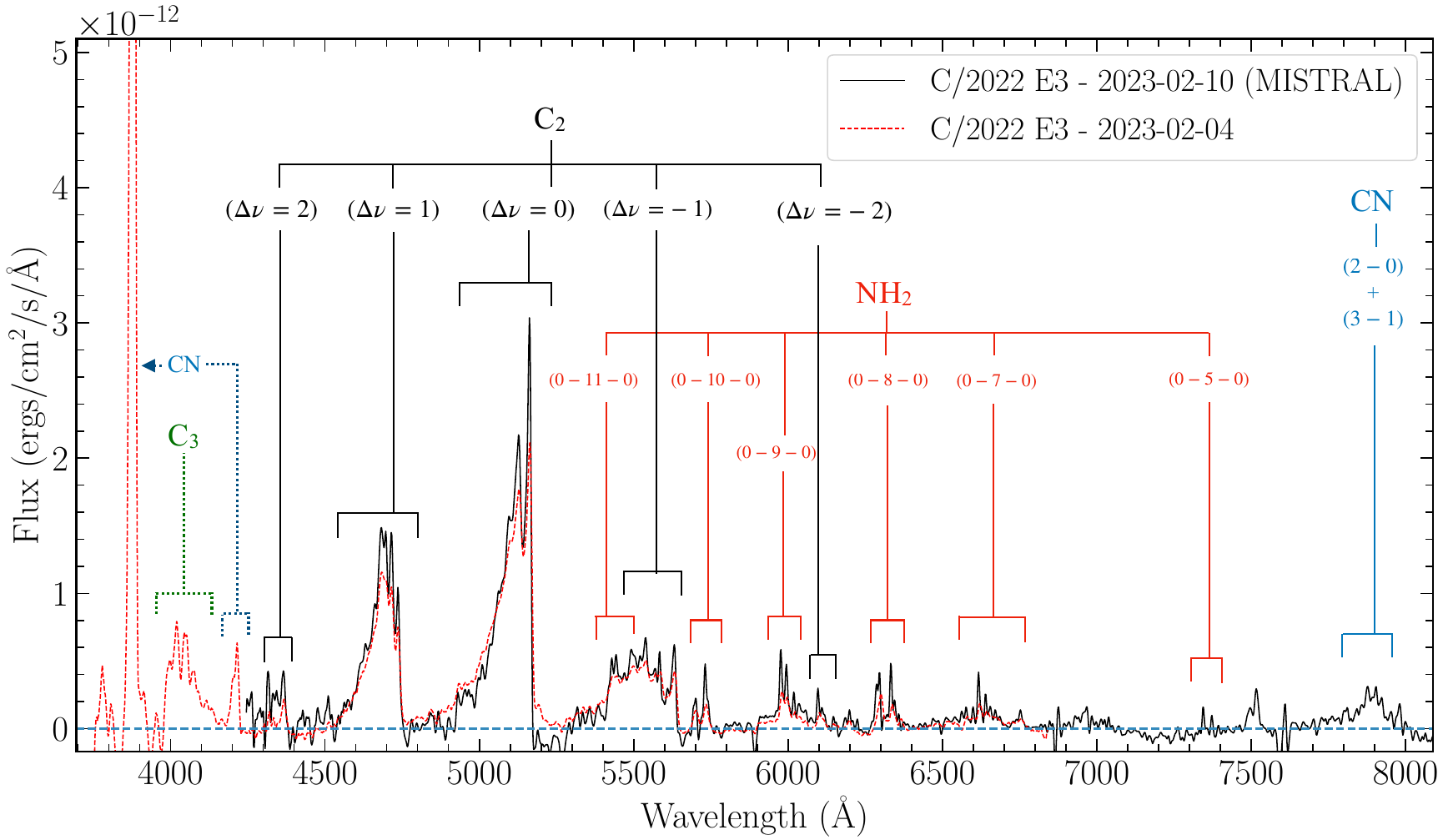}
    \caption[]{Black line: optical spectrum of comet C/2022 E3 (ZTF) observed using the MISTRAL spectrograph camera on the OHP telescope. The spectra, obtained from
the 240s frame, was extracted using a 29.2arcsec$\times$1.93arcsec aperture with the comet at the centre. Red dashed-line: spectrum obtained from the HCT telescope (private communication). }
    \label{fig:FigureC2022E33}
\end{figure}

The bright comet C/2022 E3 (ZTF) was observed in manual tracking mode for two exposure times, 120s and 240s, using the MISTRAL spectrograph
camera on the 1.93 m OHP telescope on  February 10th, 2023, when the comet was at a distance of 1.20 AU from the Sun and 0.39 AU from the Earth. We used the
blue mode covering the $\sim$4200-8000~$\AA$ range at resolution $\sim$750 with a 1.9arcsec width slit.
All the acquired frames for different exposures were bias subtracted and then flat fielded using
the spectral flat lamp present in the instrument (Tungsten). The cosmic rays were removed using the LACOSMIC package \citep{VanDokkum2001}.
The reduced and extracted 1D spectrum is shown in Fig.\ref{fig:FigureC2022E33}.

Both the 120s and 240s spectra, even if slightly trailed, were used to compute the production rates of C2 ($\Delta \nu $ = 0) emission
band using a single aperture (200 pixel aperture) with the comet at the centre, as mentioned in \cite{Aravind}.
The 120s spectra that were better guided with a narrower dust continuum were used to extract spectra with apertures of equal widths moving away from the photo
centre to compute spatial profile of the column density and hence the production rates with the help of \cite{1957BSRSL..43..740H} modelling. 
The spectrophotometric standard, Hiltner600, observed on the same night using the same configuration was used to produce the sensitivity curve of
the instrument and then flux calibrate the comet spectra.
The Solar analog standard star HD19445 was observed on the same night to remove the dust reflected solar continuum present
in the optical spectra.
Even though, ideally, a separate sky frame of similar exposure is required to remove the sky background from the comet frames, we used the sky from
a point source observed with a similar exposure time on the same night to remove the sky lines.

Several usual cometary emission bands were detected from C2, NH2 and CN radicals.
The production rate of (2.61+/-0.04)E26 and (2.51+/-0.07)E26 for C2 ($\Delta \nu $ = 0) were computed for the 240s frame and 120s respectively and
are in good agreement with the values reported for the same comet from the TRAPPIST telescopes (\citep{2022ATel15822....1J}; ATel 15822). The
Af$\rho$, a proxy to the dust production \citep{1984AJ.....89..579A} computed for the green continuum narrow band filter
\citep{2000Icar..147..180F}, from
spectroscopic data using techniques as mentioned in \citep{Aravind22}, was found to be around 5100$\pm$80 cm, which is again in agreement with the
values reported from TRAPPIST for the comet at similar observational epochs.

As a comparison, we also show in Fig.\ref{fig:FigureC2022E33} the spectrum obtained for comet C/2022 E3 from the 2 meters class Himalayan Chandra telescope (HCT, private communication). This telescope is using the Hanle Faint Object Spectrograph and Camera (HFOSC) and covers a wavelength range of 3750-6850~$\AA$. The comparison shows the nice resolution of MISTRAL showing the emission lines in a sharper way. MISTRAL is also going deeper in the red but 
is also missing the blue part of the 
spectrum with the bright CN line,  well visible in the HCT spectrum. This, e.g., justify the new
lens we are developping, more efficient toward the blue.

With the help of these OHP observations, the MISTRAL spectrograph camera on the OHP telescope proves to be reliable for cometary observations for a
wavelength range of 4200~$\AA$ - 8000~$\AA$. With proper differential tracking and required frames acquired in an orderly manner, longer exposure observation
of fainter and more distant comets could be used to effectively analyse the column density profiles of the various emission bands marked in the figure,
which can further be used to compute their production rates.

\section{Main targets of MISTRAL: the GRB follow-up program}
\label{GRBR}

The MISTRAL instrument is well-suited for studying the GRB physics (see e.g. \cite{2023sf2a.conf..503S}) as it combines both a low-resolution spectroscopic mode (with a
blue and red range) sufficient to infer their redshift using absorption lines in the afterglow emission, and  an imaging mode that allows to
characterize their optical and NIR temporal evolution.
The French GRB community has proposed a dedicated MISTRAL ToO program in order to characterize the afterglow emission of the well-localized bursts
detected by the BAT and XRT  high-energy instruments on board the \textit{Neil Gehrels Swift} space observatory \citep{2004ApJ...611.1005G}. This program is  preparatory
for the future exploitation of the Sino-French \textit{SVOM} mission expected to be launched  in June 2024 \citep{2016arXiv161006892W}, and has been conducted since March 2022. 

We have presently strict trigger criteria  to avoid disturbing too much the other observing programs, but in 
 practice, they  can be relaxed for important  events (e.g. following information reported
by other teams via Gamma-ray Coordination Network (GCN) circulars\footnote{\url{https://gcn.nasa.gov/circulars/}}) and will need to be revised once SVOM will be in operations. So far, we have observed 21 GRBs as shown in Table \ref{tab:GRB_summary}.

Most of the observed GRBs in Table \ref{tab:GRB_summary} were day-time triggers explaining the relatively long delay between trigger and MISTRAL observations. This is due to the pointing strategy of the Swift spacecraft that can detect bursts near the Sun direction. On the contrary, SVOM has an optimized anti-solar pointing strategy which will result in detecting GRBs only during the night time on Earth and thus will enhance the rates of fast follow-up at the OHP site. Few GRBs were observed with very long latency (GRB 221009A and AT2023lcr)  because newly received information (updated position and flux measurements) allowed us to identify these events as interesting targets. For now, the only (Swift) night-time trigger was GRB 240218A, which we were able to observe 30 min after the alert. We could in principle be faster, but the telescope rules impose to finish the exposure of the ongoing regular observing program if less than 30 minutes are remaining. This therefore mechanically limits the reaction time to $\sim$30minutes.

\begin{table*}[!]
    \centering
     \caption{The list of the GRBs followed up at OHP using the MISTRAL instrument mounted on the 193cm telescope during the semesters 2022A, 2022B, 2023A and 2023B.}
   \begin{tabular}{|c|c|c|c|c|c|c|c|c|c|c|}
    \hline
        GRB & mode & setting & Exp time &detection & mag &  $\mathrm{T_{obs}-T_{GRB}}$ & MISTRAL & GCN Circ. \\    
        name &  &  & sec &  &  & hours & redshift &   \\
        \hline\hline
         220623A & imaging & red & 3300 & No & i$\ge$21.5 &15.2 & -- & \cite{GCN220613A} \\\hline
         220708B & imaging & blue & 3600 & No & r$\ge$22.3 &19.5 & -- &\cite{GCN220708B} \\\hline
         220711B & imaging & blue & 3000 & No & r$\ge$21.3 &3.1 & -- & -- \\\hline
         221009A & imaging & blue & 1800 & Yes & i=20.09 &78 & -- &\cite{GCN221009A} \\\hline
         230205A & imaging & blue & 640 & No & r$\ge$20.95 &17.8 & -- &\cite{GCN230205A} \\\hline
         230328B & imaging & blue & 3000 & Yes & Rc=20.03&4.5 & -- &\cite{GCN230328B} \\\hline
         230409B & imaging & blue & 1800 & No & i$\ge$20.6 &22.3 & -- & \cite{GCN230409B} \\\hline
        \multirow{2}{*}{230427A} & imaging & \multirow{2}{*}{blue} & 4920 & \multirow{2}{*}{No} & r$\ge$22.3 & \multirow{2}{*}{11.3} &\multirow{2}{*}{--} &\cite{GCN230427Aa}\\
         & spectro & & & & & & &\cite{GCN230427Ab}\\\hline
         \multirow{2}{*}{230506C} & imaging & \multirow{2}{*}{blue} & \multirow{2}{*}{1800} & \multirow{2}{*}{Yes} & r=20.9 & 3.8 & & \cite{GCN230506Cb}\\
         &  spectro & & & & & 4.4 & 3.7 $\leq$ z $\leq$ 4 & \cite{GCN230506Ca}\\\hline
         230510A & imaging & blue & 4200 & Yes & r=21.8 &13.2 & -- & \cite{GCN230510A} \\\hline
         AT2023lcr & imaging & blue & 4560 & Yes & r=20.84 &44.4 & -- &\cite{GCNAT2023lcr} \\\hline
         230723B & imaging & blue & 3420 & No & r$\ge$21. &10.0 & -- &\cite{GCN230723B} \\\hline
         230805B & imaging & blue & 3000 & No & r$\ge$21.5 &9.7 & -- & \cite{GCN230805B} \\\hline
         \multirow{5}{*}{230812B} & \multirow{5}{*}{imaging} & \multirow{5}{*}{blue} & 1260 & \multirow{5}{*}{Yes} & r=20.32 &25 & \multirow{5}{*}{--} &\cite{GCN230812Ba} \\
         & & & 1200 & & r=22.36 &120 & -- &\cite{GCN230812Bb}\\
         & & & 2400 & & r=22.37 &145 & -- &\cite{GCN230812Bb}\\
         & & & 5500 & & r=22.45 &169 & -- &\cite{GCN230812Bb}\\
         & & & 7200 & & r=22.37 &241 & -- & \cite{GCN230812Bc}\\\hline
          \multirow{2}{*}{AT2023sva} & imaging & \multirow{2}{*}{blue} & 900 & \multirow{2}{*}{Yes} & r=20.96 &\multirow{2}{*}{--} & & \multirow{2}{*}{\cite{2023GCN.34744....1P}} \\
                   & spectro &   &  5400 &     &   &  & z $< 3.5$ & \\\hline
         231205B & imaging & blue & 4620 & No & r$\ge$20.65 &5.2 & -- & \cite{2023GCN.35286....1A} \\\hline
         231215A & imaging & blue & 3000 & Yes & r=21.32 &7.4 & -- & \cite{2023GCN.35367....1T} \\\hline
        231216A & imaging & blue & 3000 & No & r$\ge$22.2 &3.0 & -- & \cite{2023GCN.35384....1B} \\\hline
        240204A & imaging & blue & 16200 & Yes & r=21.82 &4.1 & -- & Adami et al. (in press) \\\hline
        240209A & imaging & blue & 1800 & Yes & r=20.06  & 4.9 & -- & Turpin et al. (in press) \\\hline
        240218A & imaging  & blue & 3600 & No & r$\ge$23.0 &0.5 & -- & Adami et al. (in press) \\\hline
   
    \end{tabular}
    \label{tab:GRB_summary}
\end{table*}

\subsection{Data analysis}
For each GRB MISTRAL ToO, we start with a typical sequence of r-band images in order to detect the source and
then to accurately place the slit at the right GRB location. If the GRB remains too faint or undetected after these first images we keep observing in
imaging mode to reach a higher sensitivity. If the GRB is detected and bright enough as compared to the MISTRAL spectroscopic limiting magnitude (r$<19-20$), we perform
spectroscopic observation for about 1 hour to obtain a good S/N s6). Table B.1: It is recommended to draw the profiles of the filters followed this table.
pectrum of the source.
Both the photometric and spectroscopic data are reduced immediately in order to communicate our results in real time to the
scientific community.

\subsection{Data reduction}
We first pre-process the MISTRAL raw images with the appropriate bias and flat field calibration images. The single exposure images are then
astrometrically solved and coadded to be analyzed. The GRB positions were known at the arcsecond level of accuracy which allows us to directly
perform forced photometry. To do so, we used the Simple Transient Detection Pipeline, {\sc STDPipe}, \citep{stdpipe}. Our photometric analysis
followed the steps described below (and in the git documentation\footnote{\url{https://stdpipe.readthedocs.io/en/latest/index.html}} of the
{\sc STDPipe} project):
\begin{enumerate}
    \item We mask the saturated stars, the cosmic ray artifacts and the bad columns of pixels
    \item We extract the point-like sources (S/N$>$3) using {\sc SExtractor} \citep{Bertin96}.
    \item We calibrate the photometry of the sources using the {\sc photutils}\footnote{\url{https://github.com/astropy/photutils}} Astropy
      package \citep{photutils}. We use the PanSTARRS DR1 catalog \citep{Chambers16} as a reference, a crossmatch radius of 2 arcseconds to
      associate the cataloged sources to our {\sc SExtractor} ones, and a flux aperture radius $\mathrm{\sim 1.5\times FWHM}$ (the mean of
      the full width at half maximum of the image sources). The local background was measured in an annulus of [3-5]$\mathrm{\times FWHM}$
    \item To reduce the flux contamination from nearby sources or the GRB host galaxy (if visible in the images), we may perform a difference
      image analysis using the {\sc hotpants} code \citep{hotpants} and the the CDS {\sc HiPS2FITS} service \citep{hips2fits} 
    \item If the source is detected in the coadded image, we perform the forced aperture photometry at the GRB localization using the same
      aperture configuration as during the calibration steps
    \item If the source is not detected, we estimate the upper limit from the local background for a source that would have given a positive
      detection at a given S/N
\end{enumerate}

For spectroscopy, we used the night-time dedicated MISTRAL spectroscopic 
pipeline\footnote{\url{http://www.obs-hp.fr/guide/mistral/MISTRAL_spectrograph_camera.shtml#H5}} which is based on the 
ASPIRED \citep{2023AJ....166...13L} software. 

\subsection{Results}
For 9/21 ($\sim 45\%$) of our follow-up campaigns, we were able to clearly identify and measure the brightness of the GRB afterglows. Note that among them, we detected AT2023sva \citep{2023GCN.34744....1P}, a GRB afterglow without any high-energy trigger that was initially discovered in ZTF survey data from blind searches for fast optical transients \citep{2023GCN.34730....1V}. We also covered several observational epochs both for the brightest GRB of all time (the "\textit{BOAT}"), GRB 221009A, and for the third
brightest Fermi-GBM burst, GRB 230812B \citep[OHP data reported in ][]{2023arXiv231014310H}. Section \ref{BOAT} describes in more detail our photometric results for GRB 221009A. 

\begin{figure}
    \centering
    \includegraphics[width=\columnwidth]{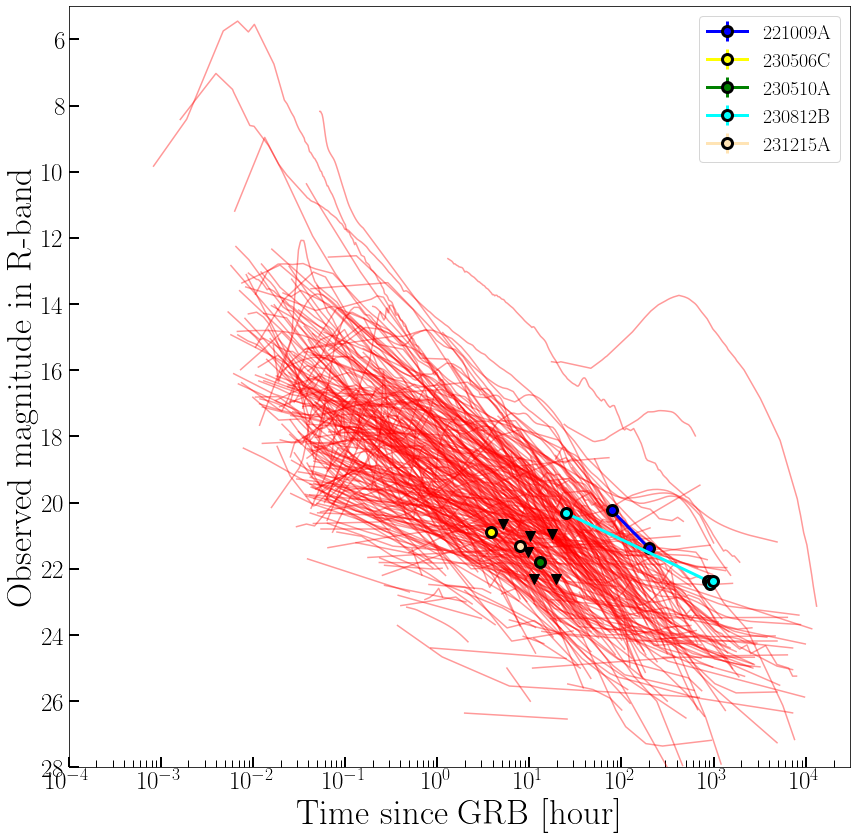}
    \caption{Comparison of the MISTRAL GRB afterglow detections (colored circles) and upper limits (black triangles) with respect to the known population of GRB afterglows (1997-2023: not a complete sample)}
    \label{fig:grb_plot_comparison}
\end{figure}

\begin{figure}
    \centering
    \includegraphics[width=\columnwidth]{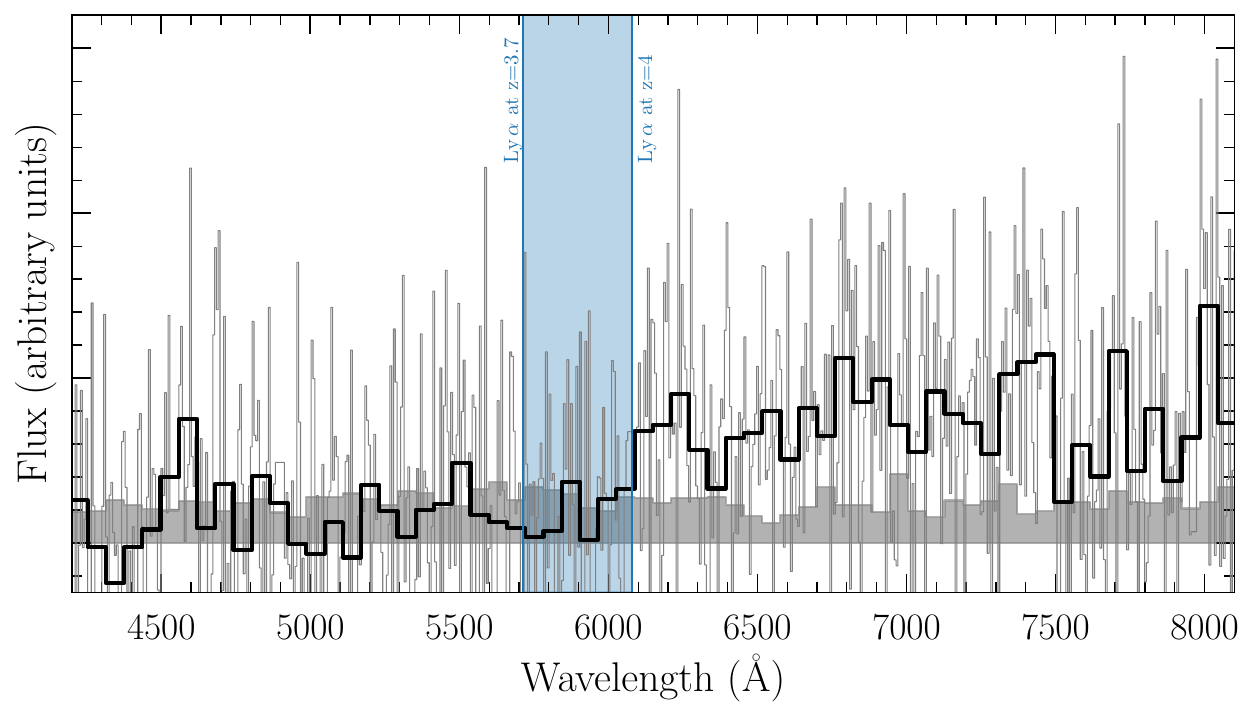}
    \caption{Spectrum of GRB230506C taken by MISTRAL (30min exposure). The thin black curve represents the full spectrum while the thick black curve represents the spectrum rebinned by 20 pixels; the gray shaded area represents the error spectrum. The center of the expected Ly$\,\alpha$ positions for $z=3.7$ and $z=4$ are shown as the blue vertical lines ; the blue shaded area represents the possible values of the Ly$\,\alpha$ centroid, assuming the drop at 6100~$\AA$ is due to Lyman break.}
    \label{fig:GRB230506C_1D}
\end{figure}

In addition, we were
able to trigger four times the spectroscopic mode of MISTRAL to characterise the transient or to infer the redshift of the potential host galaxy for GRB 230427A \citep{GCN230427Ab}, for GRB 230506C \citep{GCN230506Cb} and for the orphan afterglow AT2023sva \citep{2023GCN.34744....1P}. For the GRB 230506C, we suggest a redshift of 3.7 $\leq$ z $\leq$ 4, the only redshift measurement reported for
this burst, which also represents the most distant redshift measured at OHP so far. The spectrum of GRB\,230506C showcases the capabilities of MISTRAL pushed to its limits, securing a tentative spectroscopic redshift measurement for a source at magnitude r$\sim$20.9 with a 30min total exposure time (blue dispersor). The redshift was determined taking advantage of a break present around $6000~\AA$ that was interpreted as the Lyman-$\alpha$ break (see Fig.\,\ref{fig:GRB230506C_1D}). For the 12/21
non-detected GRBs, we were able to put interesting constraints on their optical flux evolution as shown in Figure \ref{fig:grb_plot_comparison} in the r-band.

\subsection{GRB 221009A: the "BOAT"}
\label{BOAT}

\begin{figure}[h!]
    \centering
    \includegraphics[width=\columnwidth]{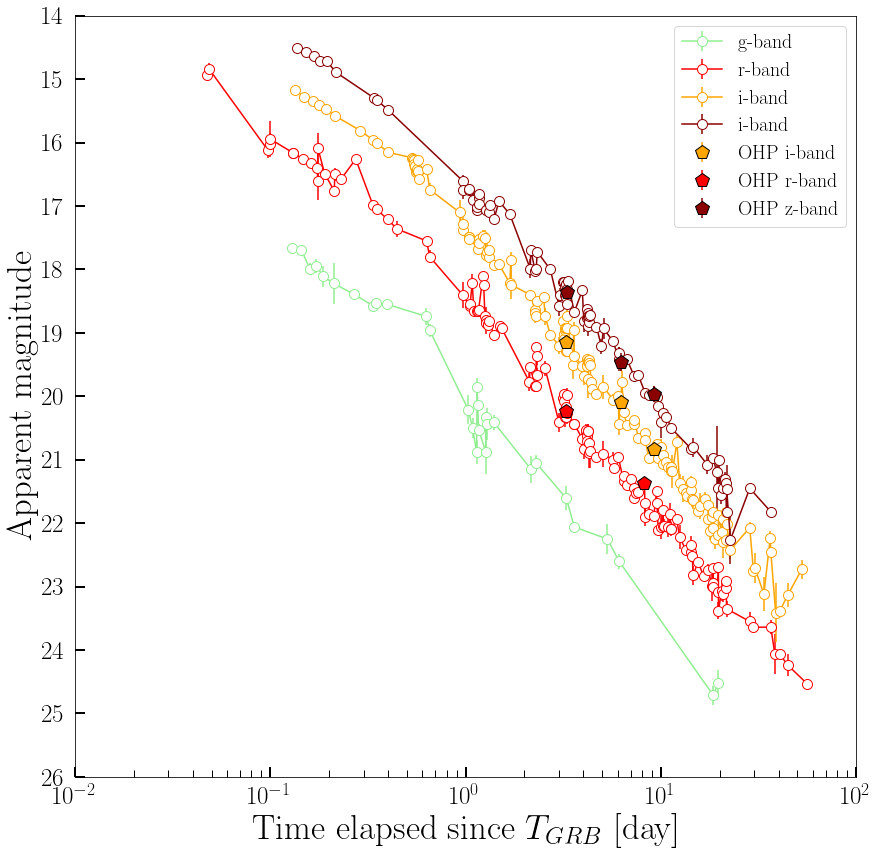}
    \caption{The optical and NIR afterglow lightcurve of GRB 221009A. The OHP observations (T193/MISTRAL and T120) are shown as colored
      pentagons. The literature data are collected from \citep{2023ApJ...946L..22F,2023ApJ...946L..23L}.
      }
    \label{fig:lc221009A}
\end{figure}

GRB 221009A was triggered by the Gamma-ray Burst Monitor (GBM) instrument onboard \textit{Fermi} \citep{2022GCN.32636....1V,2023ApJ...952L..42L}. It has been then detected during days to
weeks at all wavelengths including at TeV energies \citep{2023ApJ...946L..22F,2023ApJ...946L..23L,2023ApJ...946L..24W,2023ApJ...946L..27A,2023ApJ...946L..28L,2023arXiv230301203A,2023ApJ...948L..12K,2023SciA....9J2778C}. It was rapidly established that the properties of this burst were extraordinary given its
intense luminosity and close distance \citep[i.e. z=0.151][]{2023arXiv230207891M}. Many aspects of the GRB 221009A physics regarding the nature of the GRB relativistic jet and its angular structure, the
efficiency of the jet's kinetic to radiation conversion mechanisms and the afterglow radiation
model \citep{2023MNRAS.524L..78G,2023arXiv231012856Z,2023SciA....9I1405O,2023arXiv231015886R,2023arXiv231114180Z,2023ApJ...957L..32D}, or the Lorentz invariance principle \citep{2022arXiv221202436V,2023ApJ...942L..21F,2023arXiv230803031P,2023arXiv231209079Y}, have been highly debated in the scientific community. These discussions will continue in the upcoming
months/years as such an event is believed to occur only once every 10,000 years \citep{2023ApJ...946L..31B} and is therefore a rare opportunity to study the GRB physics with great details. In this very active context,
we have performed two epochs of photometric observations at about 6 and 10 days after the \textit{Fermi}-GBM trigger time, complemented with
three epochs with the 120cm telescope at OHP \citep{GCN221009A}. In Table \ref{tab:photometry_grb221009A}, we summarized our photometric results and showed them
in the context of the worldwide follow-up campaigns in Figure \ref{fig:lc221009A}.

\begin{table}[h!]
    \caption{Optical and NIR observation of GRB 221009A with the T193/MISTRAL and T120 telescopes located at OHP. The reported magnitudes are not corrected for the significant galactic extinction E(B-V) = 1.32 mag.}    
    \centering
    \begin{tabular}{|c|c|c|c|c|c|}
    \hline
        $T-T_{GRB}$ & filter & facility & magnitude\\ 
        (days) & & & (AB)\\
        \hline\hline
        3.259 & i' & T120 & $19.14\pm 0.08$ \\
        3.273 & r' & T120 & $20.23\pm 0.09$\\
        3.296 & z' & T120 & $18.35\pm 0.08$\\
        6.214 & i' & T193/MISTRAL &$20.09\pm 0.11$\\
        6.242 & z' & T193/MISTRAL &$19.46\pm 0.14$\\
        8.235 & r' & T120 & $21.37\pm 0.10$\\
        9.220 & i' & T193/MISTRAL &$20.83\pm 0.11$\\
        9.262 & z' & T193/MISTRAL & $19.96\pm 0.13$\\ \hline
    \end{tabular}
\label{tab:photometry_grb221009A}
\end{table}

\section{Conclusions}  

The spectro-imager MISTRAL is installed at the folded  Cassegrain focus of the 1.93m telescope at Haute-Provence Observatory since 2021 and running smoothly. It covers the spectral range from 4000~$\AA$ to 10000~$\AA$ in two settings, blue or red, at an average spectral resolution of 700. At the moment, two grisms, and five broad-band filters (g', r', i', z', Y) are available, plus some narrow-band filters around the main emission lines. Room is available for more grisms and filters to be installed in the future. The total throughput has been measured around 22$\%$ at the peak efficiency of 6000~$\AA$, with the present deep-depletion CCD of 2048 pixels of 13.5 microns  each. The efficiency is dropping rapidly below 4000~$\AA$ due to the poor transmission of the actual "blue" camera objective, but this will be replaced by a custom-made new  objective which will cover the full spectral range, avoiding time-consuming changeovers. A limiting magnitude of r$\sim$19 can be obtained in one hour of spectroscopy, while limiting magnitudes in the range of 20-21 are achieved in imaging mode in 10-20mn exposures, depending on the exact filter and local seeing conditions. A specific interface is available, showing the nature and position of all optical elements, making the instrument very users friendly. It can be put into operations in less than 15mn (change-over from the other, permanently mounted instrument, Sophie), as required by the scientific program. On-sky tests with various types of objects have shown the operability and versatility of Mistral: we give a few examples of emission-line stars, novae, young stellar objects or galaxies, with  particular emphasis on the follow-up of GRB's,  which are likely to become a major source of targets with the launch of the SVOM satellite in 2024.

\begin{acknowledgements}

 Authors acknowledge the efficient support of  the OHP night operators, in particular J. Balcaen, Y. Degot-Longhi, S. Favard, and J.P. Troncin.

Based on observations made at Observatoire de Haute Provence (CNRS), France, with MISTRAL on the T193 telescope, with AURELIE on the T152 telescope and with the CCD camera of the T120 telescope. 

We warmly thank Marco Lam for his contribution to the MISTRAL spectroscopic  reduction software. We also acknowledge very useful discussions, during the conception phase, with the SPRAT team, in particular Iain Steele and Andrzej Piascik. 
This research has made use of the MISTRAL database, operated at CeSAM (LAM), Marseille, France.

This work received support from the French government under the France2030 investment plan, as part of the initiative
   d’Excellence d’Aix-Marseille Université- A*MIDEX (AMX-19-IET-008). We were also supported by the IPhU Graduate School program at Aix-Marseille
   University. EJ is a FNRS Senior Research Associate.  TRAPPIST
   is a project funded by the Belgian Fonds (National) de la Recherche Scientifique (F.R.S.-FNRS) under grant T.0120.21. We also acknowledge the support
   by Master Erasmus Mundus Europhotonics (599098-EPP-1-2018-1-FR-EPPKA1-JMD-MOB) financed by EACEA (European Education and Culture Executive Agency). Authors thank the CNES for financial support of the MISTRAL operations. Authors thank the former Pyth\'eas institute director, N. Thouveny, for his great contribution to the MISTRAL instrument.
Authors thanks T. Adami and J. Hornstein for their contributions to the GRB alert system and the data reduction tools. Authors thank A. DelSanti for useful discussions for the C/2022 E3 (ZTF) comet observations.

\end{acknowledgements}

\bibliography{report} % bibliography data in report.bib
\bibliographystyle{aa}

%################ Appendix ########################
\newpage

\begin{appendix}

\section{Exposure Time Calculators}
\label{ETCs}

We describe below the various parameters needed for the Exposure Time Calculators to operate, and how we obtained the results presented. ETC1 is for spectroscopy, and ETC 2 for imaging. 
We note that ETCs will be updated when the single lens described in Section \ref{SingleO} will be installed on the instrument.

\subsection{Spectroscopy}
 ETC1 requires to enter several parameters:  the choice of grating (blue/red),  the expected seeing,  the
target's V band magnitude,  the required
S/N for the expected most intense spectral line,  the nature of this line (absorption or emission), and  the physical shape of the target (point source
or extended source modelled by a
Gaussian). ETC1 is basically using a N-parameter space of these parameters and fits a model at the place
of the target in the considered space in order to give an exposure time.

 ETC1 requests also : (1) the Moon conditions in terms of Moon illumination (choice between 0.25, 0.5, 0.75 and 1) and
object angular distance to the Moon (choice between 45$^{\circ}$, 90$^{\circ}$, 135$^{\circ}$ and 180$^{\circ}$),  (2) the extended (or not) nature of the object
(3) if needed, the intrinsic FWHM of the object if modeled by a Gaussian. This is the angular size of
the object before seeing convolution. At this step, ETC1 is computing the percentage of flux arriving on the CCD after convolution of
the object's signal by the seeing and the slit (1.9arcsec wide at the moment). This percentage will be applied to the
given target magnitude.  (4) The Air Mass (using the OHP extinction curve). One can also add an additional extinction in
magnitude to take into account a potential cloud-induced extinction. 

\begin{figure}[ht]
    \centering 
    \includegraphics[width=9.18cm]{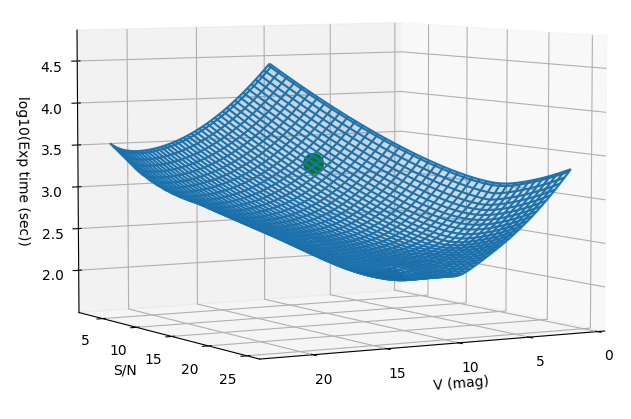}
    \caption[]{Example of ETC 3D sheet within the (magnitude, S/N, exposure time) space for the blue setting.}
    \label{fig:FigureETC1}
\end{figure}

ETC1 is now able to fit a model to real observations within the adapted space in order to give
an estimated exposure time. 3D views similar to those on Fig.\ref{fig:FigureETC1} are also provided to allow the
observer to manually adjust her/his exposure time estimate as a function of the location of the target in
the 3D sheet (see Fig.\ref{fig:FigureETC1}).

\subsection{Imaging}
Another exposure time calculator (ETC2) is available to give to the observer a typical exposure time for his/her targets in imaging mode.
ETC2 gives the exposure time needed to detect objects at a given magnitude with the requested S/N. 

These estimates are based on observations of several fields
(XCLASS3091 cluster of galaxies, \cite{2021A&A...652A..12K}, Abell400 cluster of galaxies, and at2021ft transient field) observed
in g’,r’, i’, z’ and Y bands under different seeings and airmasses. They were observed with different
exposure times, between 1 min and 60 minutes. They were all at relatively low galactic
latitudes, providing catalogs of several hundreds of stars. These stars were separated from galaxies
using the SDSS images available for these fields. We used simulations to degrade the shape of
these stars before re-detecting them (to take into account the seeing and object extension).
ETC2 is then using a N-parameter space of these characteristics and fit a polynomial function
(order 2) for the place of the target in order to give an exposure time.

Similarly to ETC1, the technique used to generate an exposure time typically consists in the following steps:
(1) select the filter you want to use in imaging. (2) give the expected seeing. (3) give the extended (or not) nature of the object
(4) if needed, give the intrinsic FWHM of the object if modeled by a Gaussian; this is the angular
size of the object before seeing convolution. (5) give an additional extinction in magnitudes to take into account a potential cloud-induced
extinction. (6) give the Air Mass to compute the magnitude loss due to the atmosphere (using the OHP extinction curve). (7) give the target's magnitude. (8) give the required S/N.

ETC2 is now able to fit a polynomial model on real observations within the chosen space in
order to give an estimated exposure time. 3D views similar to those on Fig.\ref{fig:FigureETC1} are also provided to
allow the observer to manually adjust exposure time estimate as a function of the location of
the target in the explored space. ETC2 will give the exposure time needed to
detect objects at the required magnitude to have S/N larger than the required one.

\section{Filter Responses}
\label{FRs}

\begin{figure}[ht]
    \centering 
    \includegraphics[width=9.18cm]{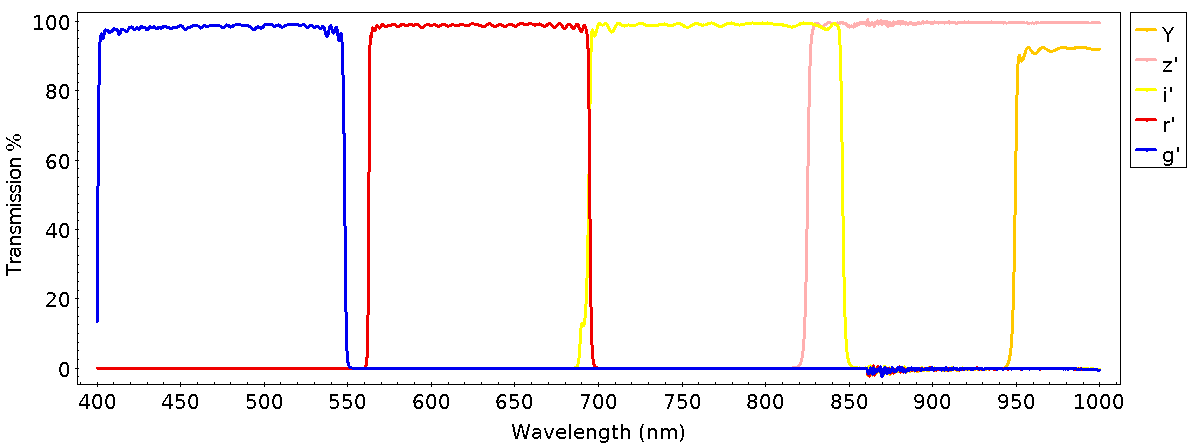}
    \includegraphics[width=9.18cm]{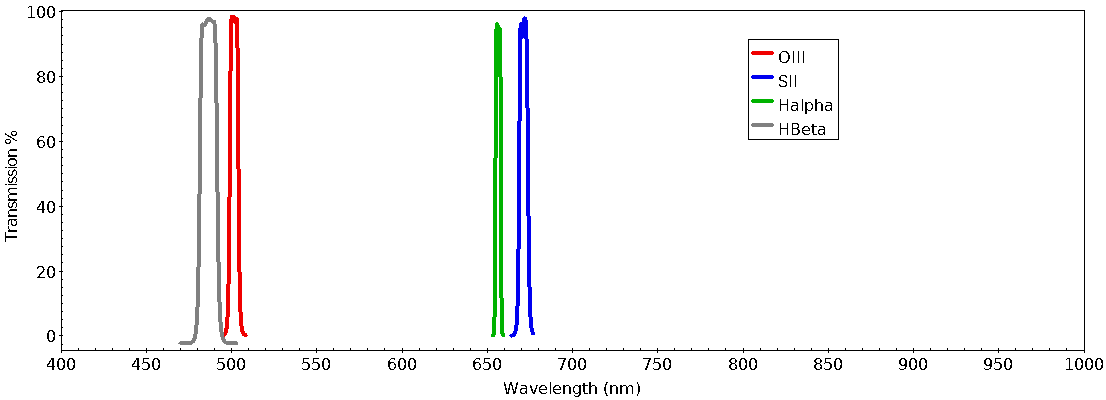}
    \caption[]{Profiles of the g', r' ,i', z', Y, H$\beta$, OIII, H$\alpha$, and SII filters described in Table \ref{tab:Tab2}.}
    \label{fig:FigureFilters}
\end{figure}

Filter responses\footnote{see \url{http://www.obs-hp.fr/guide/mistral/filter_characteristics.html}}
are summarized in Table \ref{tab:Tab2}
and shown in Fig.\ref{fig:FigureFilters}.

The MISTRAL filters transmission was measured with the Laboratoire d'Astrophysique de Marseille Perkin-Elmer spectrograph
between 4000 and 10000~$\AA$, using the PMT and InGaAs sources and low (4~$\AA$) and high (2~$\AA$) resolution. The wide range filters (g’, r’, i’, z’, Y,
passe-haut 4000-10000~$\AA$, passe-haut 6000-10000~$\AA$) were measured with a resolution of 4~$\AA$ (one measured point every 2~$\AA$ with a measured width of 4~$\AA$).
The narrow-band filters (H$\beta$, OIII, H$\alpha$, SII) were measured with same set-up between 4000 and 10000~$\AA$ and with a  better resolution of 2~$\AA$
(one measured point every $\AA$ with a measured width of 2~$\AA$) around their useful domain.

The accuracy of the measurements was evaluated  by comparing the results between two consecutive measures (without filter
repositionning between two measures) and was estimated with the SII filter measured at
high resolution. The difference between  two consecutive measures is well below the 1$\%$ level, and the shift
in wavelength is lower than 0.1~$\AA$.

\begin{table*}[ht]
\caption[]{MISTRAL filters response functions. \label{tab:Tab2}}
\centering
\begin{tabular}{ccccc}
\hline
\hline
Name                                &  Central $\lambda$    &  Maximum Transmission    &  50$\%$ low   &    50$\%$ high \\
\hline
SDSS g'                             &  $\sim$4700~$\AA$          &  $\ge$99$\%$             &  4010~$\AA$        &    5490~$\AA$ \\
SDSS r'                             &  $\sim$6300~$\AA$          &  $\ge$99$\%$             &  5630~$\AA$        &    6950~$\AA$ \\
SDSS i'                             &  $\sim$7700~$\AA$          &  $\ge$99$\%$             &  6940~$\AA$        &    8470~$\AA$ \\
SDSS z'                             &  $\sim$8400~$\AA$          &  $\ge$99$\%$             &  8250~$\AA$        &     \\
Y                                   &  $\geq$9500~$\AA$          &  $\ge$92$\%$             &  9500~$\AA$        &     \\
H$\beta$                            &  $\sim$4860~$\AA$          &  $\ge$92$\%$             &  4810~$\AA$        &    4920~$\AA$  \\           
OIII                                &  $\sim$5020~$\AA$          &  $\ge$98$\%$             &  4990~$\AA$        &    5041~$\AA$   \\          
H$\alpha$                           &  $\sim$6565~$\AA$        &  $\ge$96$\%$             &  6548~$\AA$      &    6580~$\AA$        \\     
SII                                 &  $\sim$6710~$\AA$          &  $\ge$97$\%$             &  6687~$\AA$      &    6740~$\AA$          \\   
Long Pass OD4 4000~$\AA$                 &  $\geq$4000~$\AA$          &  $\ge$98$\%$             &  4000~$\AA$        &     \\
Long Pass OD4 6000~$\AA$                 &  $\geq$6000~$\AA$          &  $\ge$98$\%$             &  5994~$\AA$      &       \\
Long Pass smooth 4000~$\AA$              &  $\geq$4000~$\AA$          &  $\ge$90$\%$             &  4000~$\AA$        &     \\
Long Pass smooth 6000~$\AA$              &  $\geq$6000~$\AA$          &  $\ge$90$\%$             &  5706~$\AA$      &       \\
\hline
\end{tabular}
\end{table*}

\section{MISTRAL spectral resolving-power estimates}
\label{VarR}

\begin{figure}[t]
\centering
\includegraphics[width=\columnwidth]{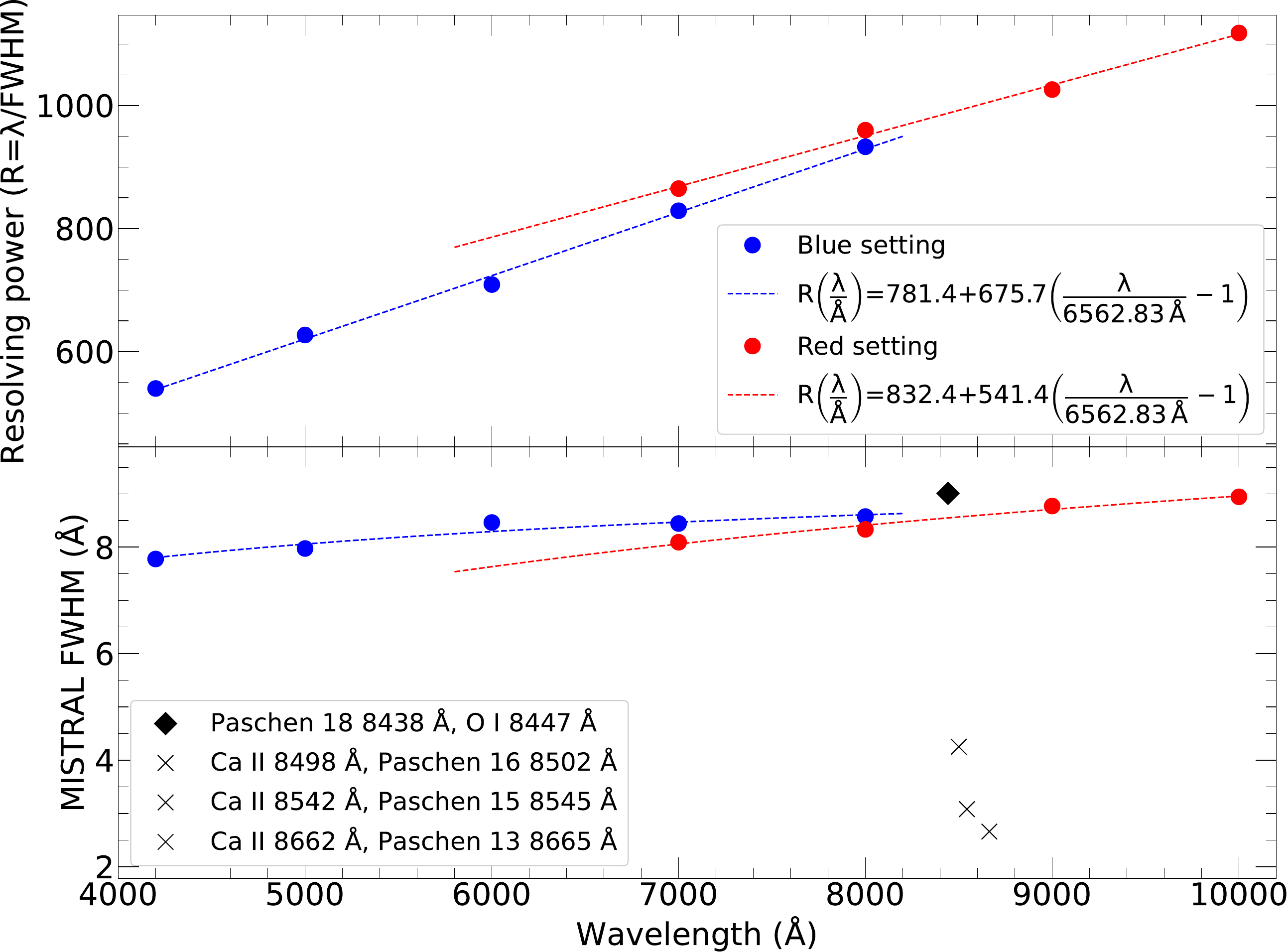}
\caption{Theoretical spectral resolving-power
of MISTRAL. 
The data for the blue and red configurations are from Table~\ref{appendix_table:R}.
{\it Top panel:} resolving power vs.\ wavelength. 
The straight lines are the linear fitting of these data.
{\it Bottom panel:} FWHM vs.\ wavelength.
The curves are computed from the linear fits.
For comparison, the wavelength separation of several close stellar emission lines
are marked with diamond and crosses 
if resolved or not, respectively, with MISTRAL.
}
\label{appendix_fig:R}
\end{figure}

The spectral resolving-power, 
$R \equiv \lambda /\textrm{FWHM}$ 
with FWHM the full-width at half maximum 
of the sharp line \citep{robertson13},  
was computed for the blue and red configurations of MISTRAL
(Table~\ref{appendix_table:R}).
The linear fitting of these data (Fig.~\ref{appendix_fig:R}) 
provides the following formulae
of $R$ vs.\ the wavelength for the blue and red configurations, 
normalised at the wavelength 
of the H$\alpha$ line:

\begin{equation}
R^\mathrm{Blue}\!\left(\frac{\lambda}{\textup{\AA}}\right) = 781.4 + 675.7 \left( \frac{\lambda}{6562.83\textup{~\AA}} -1 \right)
\end{equation}
\begin{equation}
R^\mathrm{Red}\!\left(\frac{\lambda}{\textup{\AA}}\right) = 832.4 + 541.4 \left( \frac{\lambda}{6562.83\textup{~\AA}} -1 \right)
\label{equation:Red}
\end{equation}

Therefore, the spectral resolving-power at the 
H$\alpha$ line is $\sim$7\% better in the red configuration. 

From high-resolution ($R\sim33,000$) optical spectra
of T~Tauri stars and brown dwarfs, 
it was demonstrated empirically that the full width at 10\% of the 
H$\alpha$ emission profile peak 
(hereafter FW10\%H$\alpha$) can be used as an indicator of accretion 
when it is larger than 270~km~s$^{-1}$ \citep{white03}.
To use this criteria with MISTRAL spectrum, 
the observed H$\alpha$ FWHM
must be first corrected of the MISTRAL (Gaussian) line-spread function 
to estimate the intrinsic FWHM of H$\alpha$:
$\mathrm{FWHM}_\mathrm{int}^2 = \mathrm{FWHM}_\mathrm{obs}^2 - \mathrm{FWHM}_\mathrm{MISTRAL}^2$. 
Then, using 
$\mathrm{FW10\%H\alpha}=\sqrt{\ln(10)/\ln(2)}\,\mathrm{FWHM}\,$,
we derive the velocity width corresponding to FW10\%H$\alpha$
vs.\ the observed H$\alpha$ FWHM:

\begin{equation}
  \begin{aligned}
\frac{v}{\mathrm{km\,s^{-1}}} =  & \sqrt{\frac{\ln(10)}{\ln(2)}} \times \frac{c}{R(H\alpha)} \times \\
    & \sqrt{\left(R(H\alpha)\frac{\mathrm{FWHM}_\mathrm{obs}}{6562.83} \right)^2 -1}\,,
    \label{equation:velocity}
    \end{aligned}
\end{equation}

with $c$, the speed of light.

\begin{table}[!h]
\caption{MISTRAL theoretical spectral resolving-power.}
\label{appendix_table:R}
\centering
\begin{tabular}{cc}
\hline\hline
Wavelength & $R$ \\
(\AA) & \\
\hline
\multicolumn{2}{c}{Blue configuration}\\
4200 & 540\\
5000 & 627\\
6000 & 709\\
7000 & 829\\
8000 & 933\\
\hline
\multicolumn{2}{c}{Red configuration}\\
7000 & 865\\
8000 & 960\\
9000 & 1026\\
10000 & 1118\\
\hline
\end{tabular}
\end{table}

\section{Spectrograph sensitivity
and telluric corrections using Feige~15
in the red configuration of MISTRAL
}
\label{appendix:telluric}

\begin{figure*}[t]
\centering
\includegraphics[width=2\columnwidth,trim={0 0 0 0.5cm},clip]{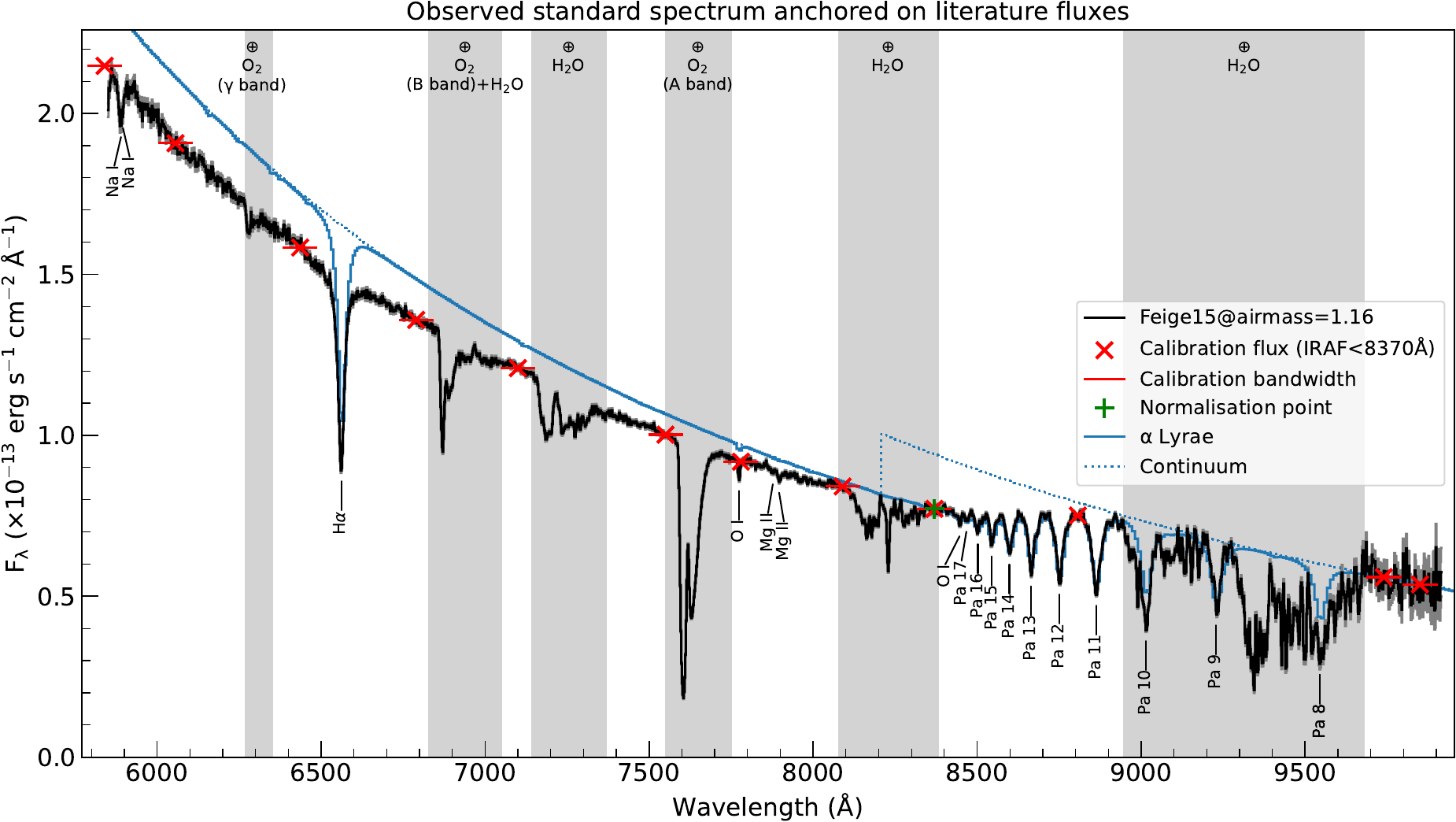}
\caption{MISTRAL red configuration spectrum of 
the spectrophotometric standard Feige~15, obtained 
with an exposure of 600~s
on the night of 2021 Dec.~8 (no telluric correction). 
The flux error-bars are plotted in gray.
The vertical light-gray stripes indicate the atmospheric 
absorption bands of water and molecular oxygen \citep{lu21}. 
The reference flux and bandwidth wavelengths are marked in red 
(\citealt{stone77}, shortwards of 8370~\AA).
The solid and dotted blue lines are 
the $\alpha$~Lyrae (Vega) 
CALSPEC
spectrum and continuum 
\citep{bohlin14}, 
respectively,
normalised at the wavelength and flux marked by the green plus.}
\label{appendix_fig:feige15}
\end{figure*}

The spectrophotometric standard Feige~15 is 
a faint, blue star of the galactic halo \citep{feige58}.
The determination of its spectral type 
has evolved from A1~V \citep{sargent68} to sdA0IV:He1 
\citep{drilling13}.
Its spectral energy distribution (SED) is 
only reported
from 3200 to 8370~\AA\ \citep{stone77}, 
using band withs of 49~\AA\ and 98~\AA\ in the 
wavelength ranges 3200--5263 and 5263--8370~\AA, 
respectively \citep{stone74}. 
Therefore, Feige~15 is routinely used 
for the spectral calibration of MISTRAL 
in the blue configuration; but, 
its SED must be extended
above 8370~\AA\ to be used in the red configuration.

We have adapted the {\tt get\_sensitivity} 
function of ASPIRED\footnote{\url{http://www.obs-hp.fr/guide/mistral/MISTRAL_spectrograph_camera.shtml#H5}}  \citep{2023AJ....166...13L} to use the bandwiths 
of the reference fluxes 
when determining the spectrograph sensitivity 
from the standard spectrum. 
We have written a custom python 
script to use 
the CALSPEC\footnote{\url{https://www.stsci.edu/hst/instrumentation/reference-data-for-calibration-and-tools/astronomical-catalogs/calspec}{\url{https://www.stsci.edu/hst/instrumentation/reference\-data-for-calibration-and-tools/astronomical-catalogs/calspec}}} 
spectrum of $\alpha$~Lyrae (Vega), 
normalised on the band width centered at 8370~\AA,
to extend the SED longwards.
We compute the fluxes of this A0 template within
bandwidths of 25, 100, and 100~\AA\ centered
at 8807 (mean of the Pa~11 and Pa~12 wavelenghs), 
9740, and 9850~\AA, respectively, 
outside the atmospheric absorption band of water \citep{lu21}, 
which leads to 10.679, 10.780, and 10.800~ABmag, respectively. 

Fig.~\ref{appendix_fig:feige15} shows the 
reduced spectrum of Feige~15 as observed at airmass 1.16,
where the standard fluxes match the literature 
fluxes \citep{stone77} plus these three new references 
of flux. The observed depths of the Paschen lines 
have a very good match outside the atmospheric 
absorption band of water with the A0 template, 
which can hence also be used for telluric correction 
in this atmospheric absorption band.

We estimate the standard continuum shortwards 
of 8370~\AA\ after excluding the atmospheric 
absorption bands. 
The ratio inside the atmospheric 
absorption bands between 
the standard continuum (shortwards of 8370~\AA) 
or the A0 template (longwards of 8370~\AA)
and the observed flux 
is used for the telluric correction of the spectrum
of LkH$\alpha$~324SE
(Fig.~\ref{fig:LkHA_324SE}).

\section{CCD reading modes}
\label{ReaMode}

Several CCD reading modes are available (nominal temperature of the CCD: $\sim$-90C, saturation
level: $\sim$60 000). Two modes have been extensively tested and they are the only ones offered: fast
mode (3 MHz) and slow mode (50 kHz). They are briefly described in Table \ref{tab:RM}. The fast mode is
adapted for technical operations such as telescope focus or pointing, thanks to the very short reading
time allowing e.g. real-time object focusing. For science
operations, a slow mode is offered with a nearly three times lower read-out noise.

\begin{table}[ht]
\caption[]{Characteristics of the different reading modes. \label{tab:RM}}
\centering
\begin{tabular}{ccc}
\hline
\hline
	&	Fast mode: 3 MHz	&	Slow mode: 50 kHz \\
\hline
Read Out Noise&	11 e-	     		&	4 e- \\
Reading time&	0.7 sec			&	40 sec \\
Gain in e-/ADU&	1.03			&	1.03 \\
\hline
\end{tabular}
\end{table}

\section{Instrument stability}
\label{InsStab}

One of the science goals of MISTRAL being the 
long-time follow-ups of variable objects, it is important to evaluate its stability over time.

\begin{figure}[t]
\centering
\includegraphics[angle=270,width=\columnwidth]{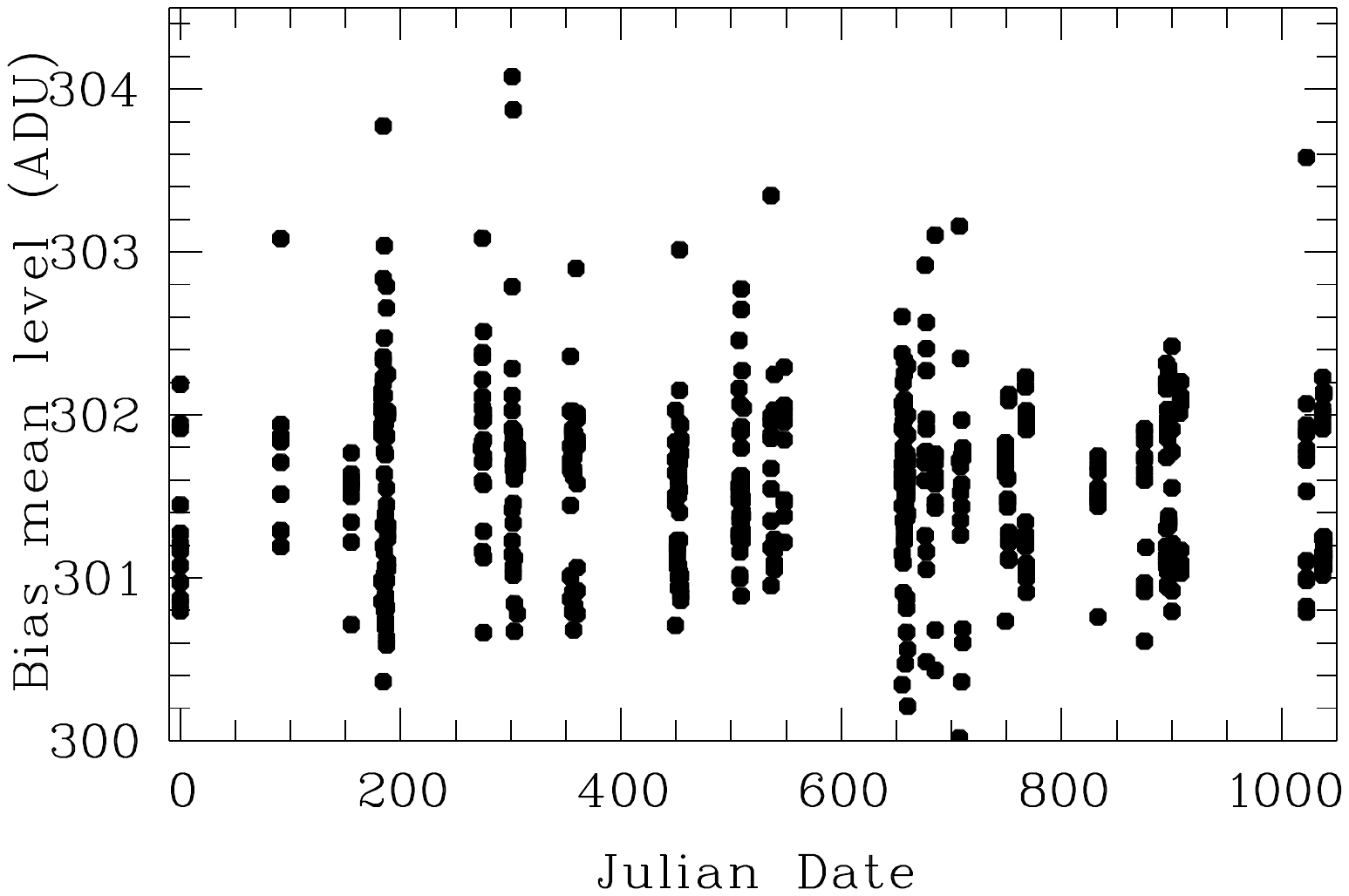}
\includegraphics[angle=270,width=\columnwidth]{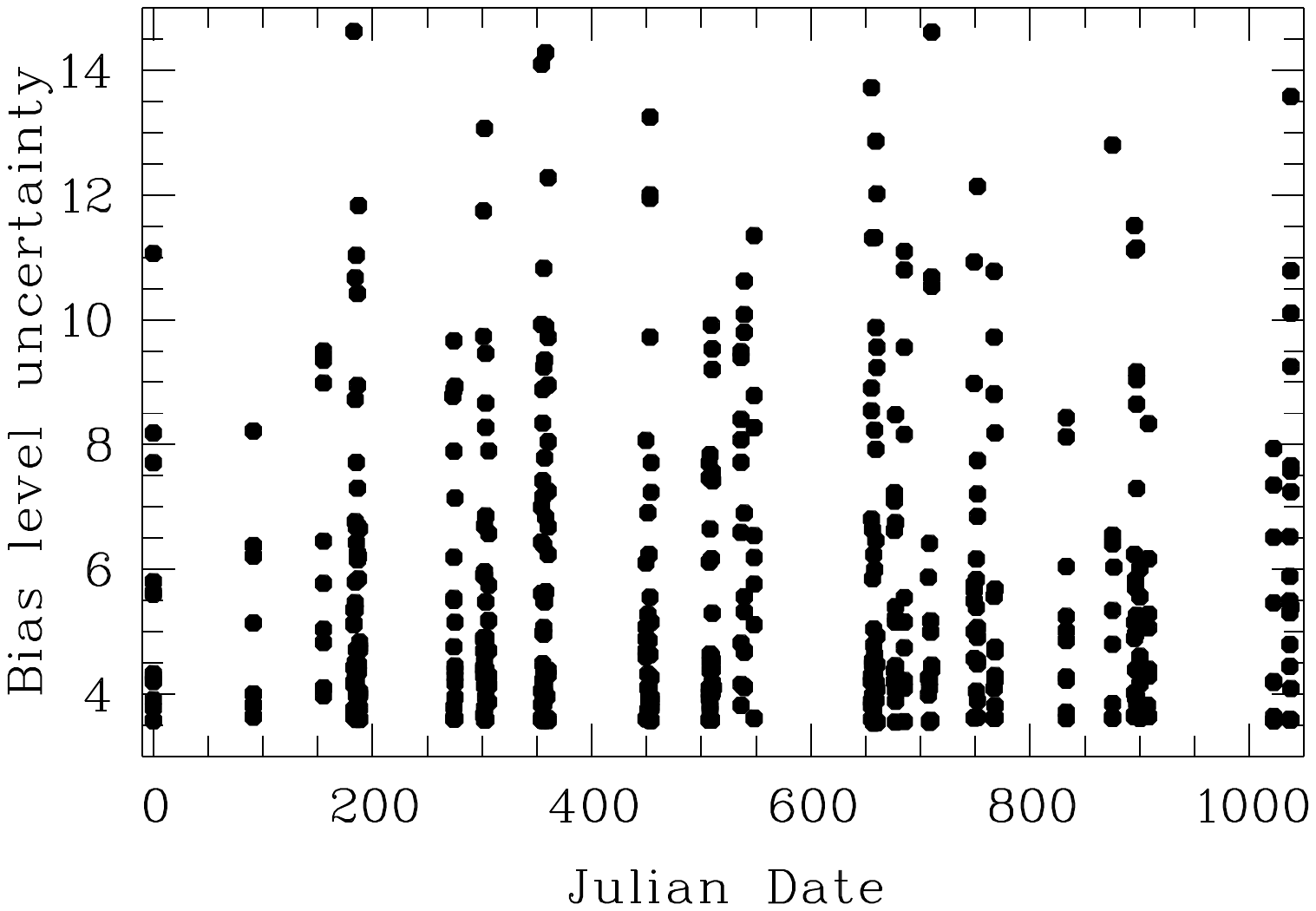}
\caption{Variation over time (starting at the Julian Date of 2459310) of the mean ADU level (top panel) and of the 1-$\sigma$ mean level uncertainty (bottom panel) of  MISTRAL's biases.
}
\label{appendix_fig:CCD1}
\end{figure}

We therefore investigated the CCD mean electronic noise level with the 760 measured bias frames over an operating period of $\sim$3 years. 
Fig.~\ref{appendix_fig:CCD1} shows the variation 
of the mean ADU level and of the 1-$\sigma$ mean level uncertainty of biases over a period of 1050 days. Both are compatible with no variation over time, with a bias level of 302$\pm$6 ADUs.

\begin{figure}[t]
\centering
\includegraphics[angle=270,width=\columnwidth]{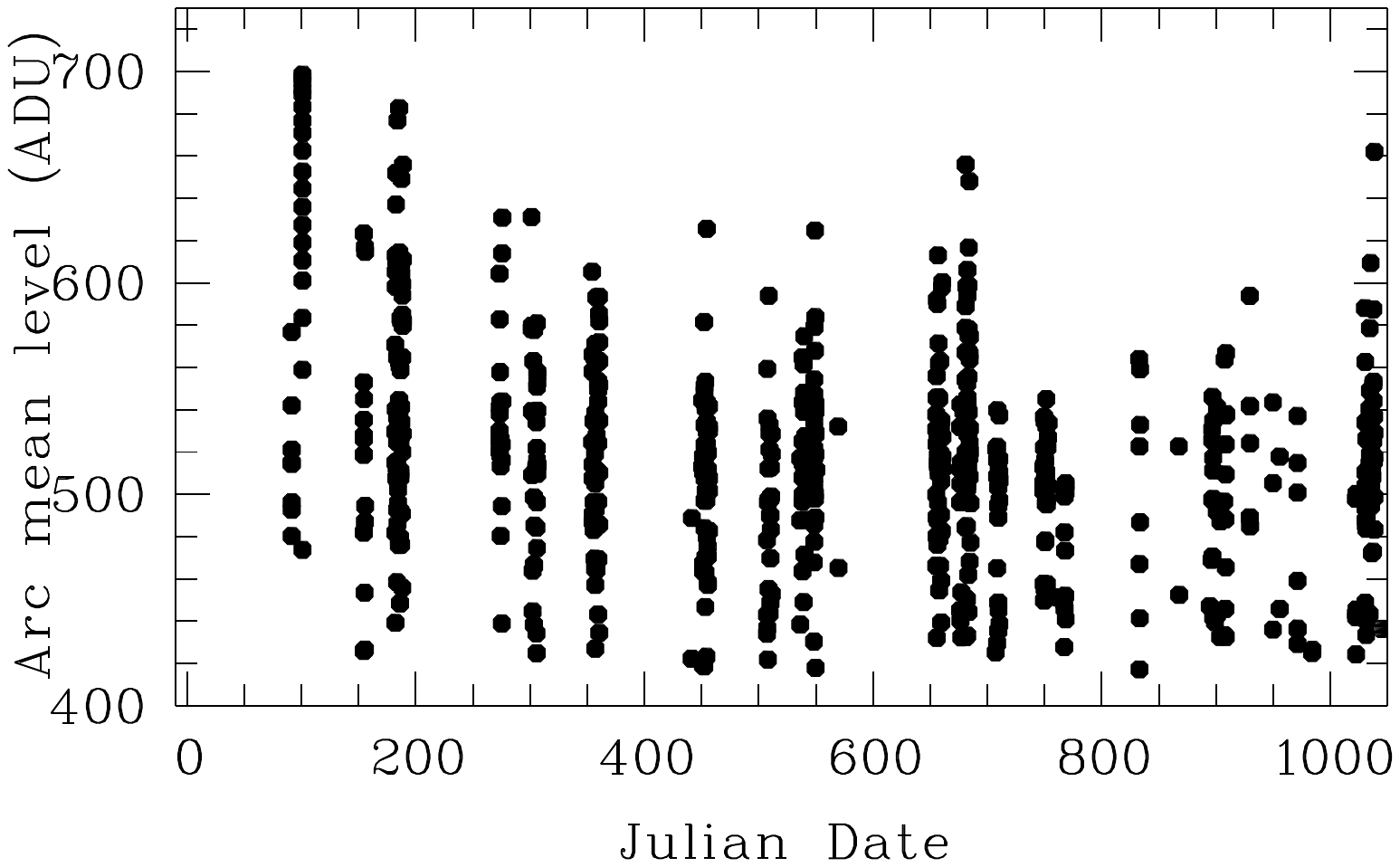}
\includegraphics[angle=270,width=\columnwidth]{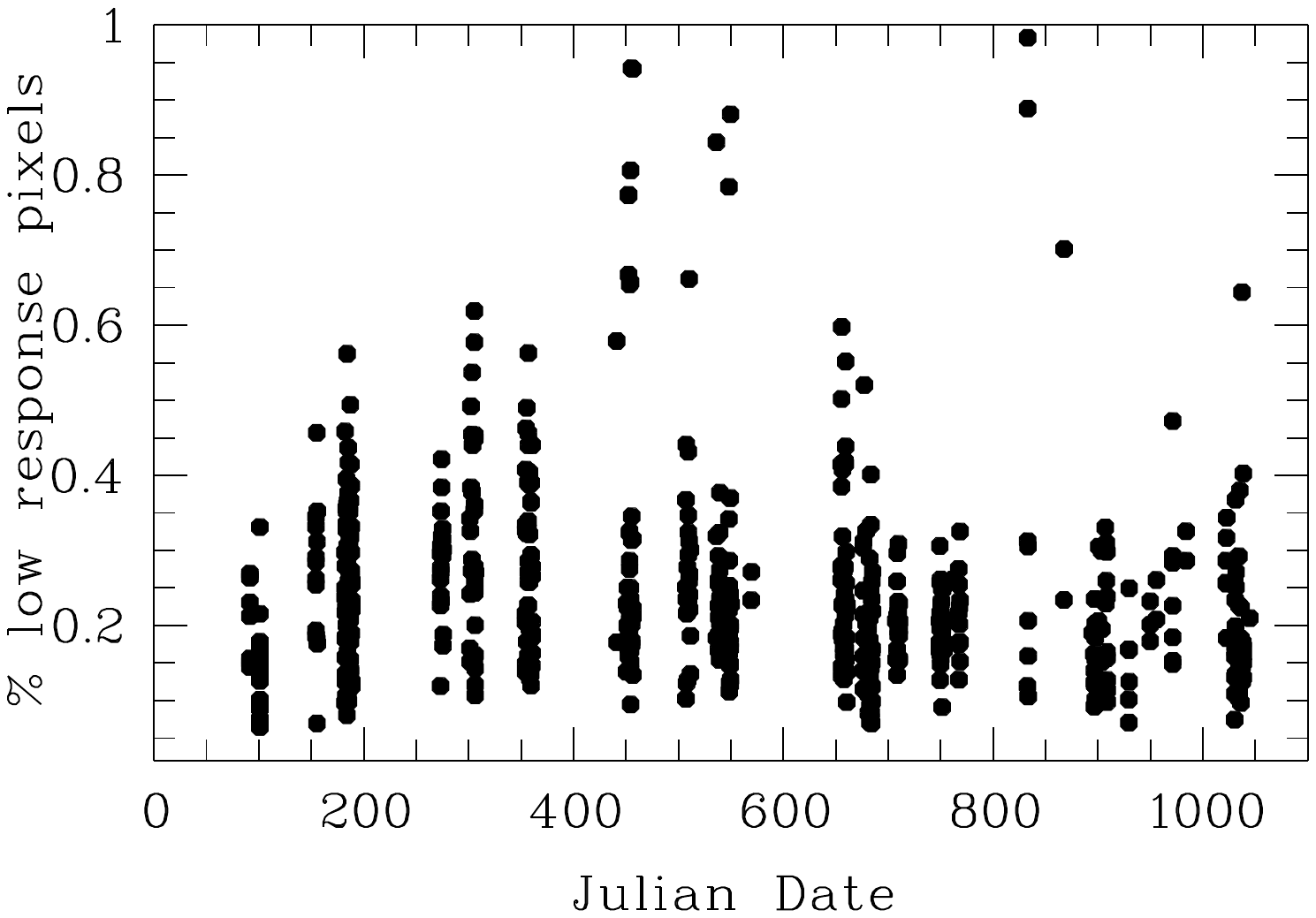}
\caption{Variation over time (starting at the Julian Date of 2459310) of the mean ADU level (top panel) of  MISTRAL's arc exposures. Bottom panel shows the percentage of potential low-response pixels over time.}
\label{appendix_fig:CCD2}
\end{figure}

We then investigated the response of the CCD when illuminated by an artificial source (to avoid sky variability and telescope primary mirror coating variations). We used for this the 917 available arc calibration frames, selecting only the 30sec exposures made with the blue dispersor.
Fig.~\ref{appendix_fig:CCD2} shows the variation of the mean flux level of these exposures (upper panel). After an initial 15$\%$-decrease of the flux during the first year probably due to the wavelength calibration lamp aging, the mean flux is remaining constant since almost two years.

We finally estimated the variation of the percentage of low-response CCD pixels. For each arc frames, we computed the number of pixels with an ADU level below 320. This corresponds to the mean bias level plus 3$\sigma$. Every pixel below this level may be a low-response pixel because still compatible with a bias level despite the 30sec exposure with the calibration lamps "on". Fig.~\ref{appendix_fig:CCD2} shows a very low and stable level of these pixels, around 0.2$\%$ of the total number of CCD pixels.

\end{appendix}

\end{document}